\documentclass{JHEP}	
\usepackage{epsfig}
\usepackage[active]{srcltx}

\newcommand{\eref}[1]{(\ref{#1})}
\newcommand{\sref}[1]{\S\ref{#1}}

\newcommand{\fref}[1]{Figure~\ref{#1}}
\newcommand{\cref}[1]{Chapter~\ref{#1}}
\newcommand{\beq}{\begin{equation}}
\newcommand{\eeq}{\end{equation}}
\newcommand{\ba}{\begin{array}}
\newcommand{\ea}{\end{array}}
\newcommand{\bcenter}{\begin{center}}
\newcommand{\ecenter}{\end{center}}

\def\IB{\relax\hbox{$\inbar\kern-.3em{\rm B}$}}
\def\IC{\relax\hbox{$\inbar\kern-.3em{\rm C}$}}
\def\ID{\relax\hbox{$\inbar\kern-.3em{\rm D}$}}
\def\IE{\relax\hbox{$\inbar\kern-.3em{\rm E}$}}
\def\IF{\relax\hbox{$\inbar\kern-.3em{\rm F}$}}
\def\IG{\relax\hbox{$\inbar\kern-.3em{\rm G}$}}
\def\IGa{\relax\hbox{${\rm I}\kern-.18em\Gamma$}}
\def\IH{\relax{\rm I\kern-.18em H}}
\def\IK{\relax{\rm I\kern-.18em K}}
\def\IL{\relax{\rm I\kern-.18em L}}
\def\IP{\relax{\rm I\kern-.18em P}}
\def\IR{\relax{\rm I\kern-.18em R}}
\def\IZ{\relax\ifmmode\mathchoice
{\hbox{\cmss Z\kern-.4em Z}}{\hbox{\cmss Z\kern-.4em Z}}
{\lower.9pt\hbox{\cmsss Z\kern-.4em Z}}
{\lower1.2pt\hbox{\cmsss Z\kern-.4em Z}}\else{\cmss Z\kern-.4em Z}\fi}
\def\II{\relax{\rm I\kern-.18em I}}

\def\sCC{{\kern 0.27em\vrule height1.45ex width0.03em depth0em
          \kern-0.30em\rm C}}
\def\C{{\mathchoice
  {\sCC}
  {\sCC}
  {\kern 0.225em \vrule height1.05ex width0.025em depth0em \kern-0.25em \rm C}
  {\kern 0.180em \vrule height0.78ex width0.02em depth0em \kern-0.2em \rm C}
        }}
\def\sHH{{\rm I\kern-.16em{}H}}
\def\H{{\mathchoice
  {\sHH}
  {\sHH}
  {\rm I\kern-.13em{}H}
  {\rm I\kern-.13em{}H} }}
\def\sNN{{\rm I\kern-.16em{}N}}
\def\N{{\mathchoice
  {\sNN}
  {\sNN}
  {\rm I\kern-.12em{}N}
  {\rm I\kern-.10em{}N} }}
\def\sPP{{\rm I\kern-.16em{}P}}
\def\P{{\mathchoice
  {\sPP}
  {\sPP}
  {\rm I\kern-.12em{}P}
  {\rm I\kern-.10em{}P} }}
\def\sQQ{{\kern 0.27em \vrule height1.45ex width0.03em depth0em
          \kern-0.30em \rm Q}}
\def\Q{{\mathchoice
        {\sQQ}
        {\sQQ}
  {\kern 0.225em \vrule height1.05ex width0.025em depth0em \kern-0.25em \rm Q}
  {\kern 0.180em \vrule height0.78ex width0.020em depth0em \kern-0.20em \rm Q}
        }}
\def\sRR{{\rm I\kern-0.16em{}R}}
\def\R{{\mathchoice
  {\sRR}
  {\sRR}
  {\rm I\kern-0.12em{}R}
  {\rm I\kern-0.10em{}R} }}
\def\sZZ{{\rm Z\kern-0.32em{}Z}}
\def\Z{{\mathchoice
  {\sZZ}
  {\sZZ} 
  {\rm Z\kern-0.3em{}Z}     
  {\rm Z\kern-0.25em{}Z} }}  
\def\ZZZ{{\rm Z\kern-0.24em{}Z}}
\def\sII{{\rm I\kern-0.16em{}I}}
\def\I{{\mathchoice
  {\sII}
  {\sII}
  {\rm I\kern-0.12em{}I}
  {\rm I\kern-0.10em{}I} }}

\def\dim{{\rm dim}}

\def\inbar{\,\vrule height1.5ex width.4pt depth0pt}
\font\cmss=cmss10 \font\cmsss=cmss10 at 7pt

\def\nn{\nonumber}

\def\smiley{\hbox{\large$\bigcirc$\hspace{-0.80em}\raise.2ex
\hbox{$\cdot\cdot$}\kern-.61em\lower.2ex\hbox{\scriptsize$\smile$}}\ }
\def\frowny{\hbox{\large$\bigcirc$\hspace{-0.80em}\raise.2ex
\hbox{$\cdot\cdot$}\kern-.635em\lower.2ex\hbox{\scriptsize$\frown$}}\ }

\def\I{{\rlap{1} \hskip 1.6pt \hbox{1}}}

\newcommand{\mat}[1]{\left( \matrix{#1} \right)}
\newcommand{\tmat}[1]{{\scriptsize \mat{#1}}}
\makeatletter
\let\hangafter\@hangfrom
\makeatother

\newcommand{\be}{\begin{equation}}
\newcommand{\ee}{\end{equation}}
\newcommand{\bea}{\begin{eqnarray}}
\newcommand{\eea}{\end{eqnarray}}
\newcommand{\bean}{\begin{eqnarray*}}
\newcommand{\eean}{\end{eqnarray*}}

\preprint{MIT-CTP-3481\\ UPR-1073-T}

\title{On Correspondences Between Toric Singularities \\
and $(p,q)$-webs}
\author{Bo Feng$^1$, Yang-Hui He$^2$ and Francis Lam$^3$\\
$^1$Institute for Advanced Study, Princeton, NJ 08540\\
$^2$Dept.~of Physics and Math/Physics RG,
	University of Pennsylvania,\\ 
	~~~~~209, S.~33rd st., Philadelphia, PA 19104.\\
$^3$Center for Theoretical Physics,
	Massachusetts Institute of Technology,\\ 
	~~~~~Cambridge, MA 02139, USA.\\~\\
\email{fengb@sns.ias.edu, yanghe@physics.upenn.edu, flam@alum.mit.edu}
}

\abstract{We study four-dimensional ${\cal N}=1$ gauge theories which
arise from D3-brane probes of toric Calabi-Yau threefolds. There are
some standing paradoxes in the literature regarding relations
among $(p,q)$-webs, toric diagrams and various phases of the gauge
theories, we resolve them by proposing and carefully distinguishing
between two kinds of $(p,q)$-webs:
{\sl toric and quiver $(p,q)$-webs}. The former has a
one to one correspondence with the toric diagram
while the latter can correspond to multiple gauge theories. 
The key reason for this ambiguity
is that a given quiver $(p,q)$-web can not capture non-chiral matter
fields in the gauge theory.
To support our claim we analyse families of theories emerging
from partial resolution of Abelian orbifolds using the Inverse
Algorithm of hep-th/0003085 as well as $(p,q)$-web techniques. We
present complex inter-relations among these theories by Higgsing,
blowups and brane splittings. We also point out subtleties involved in
the ordering of legs in the $(p,q)$ diagram.
}

\begin{document}

\newpage
\section{Introduction and Summary}
Investigating the world volume theories of D-brane probes on Calabi-Yau
manifolds is of vital phenomenological importance as well as
theoretical interest. In particular,
toric singularities have been investigated for some time
\cite{DGM,MP,Aspinwall,DM,Chris1,DD} because of the wealth of
mathematical techniques that allow a detailed study.
Amongst many other boons, progress in this direction has
taught us salient prospects on the AdS/CFT
correspondence \cite{MP,KS,Herzog:2003dj,chaos}.

A plethora of methods have been developed to 
extract field theories from these toric singularities, 
each with its virtues.
The first of these methods is to use the so-called ``Inverse
Algorithm'' developed in
\cite{Feng:2000mi,Feng:2001xr,Feng:2001bn}. This algorithm is based
on the ``Forward Algorithm'' given in \cite{DGM,Chris1,DD}. 
Another method is to use the higgsing mechanism, starting from a
parent Abelian orbifold theory 
and adjusting the FI-parameters properly in the
spacetime field theory \cite{Beasley:2001zp,Beasley,raugas,unhiggs}. 
A third method is to use
geometric engineering techniques of \cite{HIV}
to calculate the appropriate intersection
numbers and mappings among exceptional collections
\cite{Hanany:2001py,Cachazo:2001sg,Herzog:2003dj,Martijn}.
The fourth one, which we shall address in detail here,
is to use the $(p,q)$-web picture of
\cite{Aharony:1997bh,Leung:1997tw}
as a useful guide to directly
obtain the matter content and find fields which should
be higgsed down \cite{Hanany:2001py,Franco:2002ae}.
Of these four methods, the $(p,q)$-web technique is very attractive,
given the ease by  
which the matter content, i.e., quiver diagram, of the gauge theory
can be calculated  
(from intersection numbers of $(p,q)$-charges) and the immediate
identification of  
fields higgsed down by acquiring VEVs in the parent orbifold theory.

For toric singularities,
the method of $(p,q)$-webs is particularly useful due to a supposed
equivalence between the toric diagram and the $(p,q)$-web
representations \cite{Aharony:1997bh,Leung:1997tw}.
The $(p,q)$ webs provide us
with a very direct and picturesque perspective. Indeed, in
\cite{Aharony:1997bh}, 
it was noticed that the grid diagram dual to $(p,q)$-web
precisely resembles the corresponding toric diagram.
Then, in \cite{Leung:1997tw},
evidence from the geometric point of view was given to support such a
relation.
These works tell us that (at least for the case of a single interior
point in the toric diagram) 
{\em there is a one-to-one correspondence between the $(p,q)$-web and the
toric data}.

If we accept the above correspondence, the next question is to find the 
quiver theory of a given toric data or $(p,q)$-web. 
As mentioned above, the quiver diagram can be directly obtained
from the given $(p,q)$ charges.
The relation between the quiver diagram and the $(p,q)$ charges was
established in \cite{Hanany:2001py} by mirror symmetry for cones over 
del Pezzo surfaces. This begs the question:  does a one-to-one
correspondence hold for general
$(p,q)$-webs?
Along this line, \cite{Franco:2002ae} identified different
$(p,q)$-webs for different  
phases of toric duality \cite{Feng:2000mi,Feng:2001xr} 
for cones over the ample surfaces
$F_0$, $dP_2$ and $dP_3$ and established the transitions
among them by higgsing and unhiggsing mechanisms \cite{unhiggs}. 

Therefore, we run into a paradox. On the one hand, a one-to-one
correspondence would dictate 
that given a toric singularity, there should be
a unique corresponding $(p,q)$-web. On the other hand, 
different $(p,q)$-webs have been identified with different toric
(Seiberg) dual phases 
of a given toric singularity. The two situations cannot both be true
and yet each seems to have its supporting evidence.

It is the main aim of this paper to explain this puzzle. 
By careful analysis, we find that we need to distinguish two kinds of
$(p,q)$-webs.
The first one is what we shall call a 
{\bf toric  $(p,q)$-web} which has
the direct one-to-one  
correspondence with a given toric variety. This is the one
considered in \cite{Hanany:2001py,Aharony:1997bh,Leung:1997tw}.
The second one is what we shall call {\bf quiver $(p,q)$-web} 
which re-expresses
a quiver diagram in the form of a $(p,q)$-web \cite{Franco:2002ae}. 
The second concept has limitations. Firstly, the quiver $(p,q)$-web 
can be applied only for toric diagrams with 
one interior point. For those with more than one interior point,
we need to use more than one pair of $(p,q)$ charges to calculate the
quiver diagram.  Secondly, as we will show, we can use the same
``quiver $(p,q)$-web'' for several different theories and read 
out various higgsing phases; this will circumvent our problem of the
lack of one-one correspondence.
Because of these limitations, however, we
think that the first concept of the toric $(p,q)$-web
is more fundamental, although the second
does have its useful r\^{o}le.

In summary, then,
the outline of the paper is as follows. We begin with a review in
Section 2 on the
key points of $(p,q)$-webs and in particular how they represent a
four-dimensional world-volume gauge theory on D3-branes probing
singular toric Calabi-Yau varieties. We show how one obtains the web
from the toric diagram of the singularity by graph-dualisation. Such
webs we shall call ``toric $(p,q)$-webs.'' We
then go to extensive case studies of the toric partial resolutions of 
$\IC^3/\IZ_3\times \IZ_3$ in Section 3. We will reach families of
Hirzebruch, del Pezzo and
so-called pseudo del Pezzo surfaces. The quivers for these theories
are intricately inter-related and if we ignored bi-directional arrows
therein many of these theories have the same quiver even though their
toric diagrams and hence toric $(p,q)$-webs differ greatly. We
re-construct the web which gives rise to these quivers and call these
webs ``quiver $(p,q)$-webs.'' These are more useful in reading out the
higgsing information. To this we then turn in Section 4 where detailed
checks from the field theory, complete with the superpotential, are
carried out to show how these plethora of theories are related by
(un)higgsing. From the web point of view, these (un)higgsing correspond to
splitting and combining of external legs in
the web-diagrams, and we finally show how the various
theories can be reached by combining legs from the parent $(p,q)$-webs
of the $\IC^3/\IZ_3\times \IZ_3$ orbifold.

%
\section{A Brief Reminder on $(p,q)$-Webs of Branes}
We are interested in the construction of a wide class of
four-dimensional ${\cal N}=1$ gauge theories that arise on the world-volume
of D-branes probing toric singularities. A particularly visual method
of studying the quiver that arise from such theories is to use
$(p,q)$-webs. In this section, we briefly outline the techniques
involved before proceeding on to study detailed examples.
The reader is encouraged to also refer to Section 2 of
\cite{Franco:2002ae} for a very nice review on this subject in the
present context.

\subsection{The $(p,q)$-Web}
We first recall that the S-duality in type IIB, or equivalently, the
torus modular group in F-theory, allows the existence of $(p,q)$
branes. The D5-brane is assigned a charge of $(1,0)$ while the
NS5-brane has charge $(0,1)$.
The methods of \cite{HW,AH} allowed \cite{Aharony:1997bh,Leung:1997tw}
to construct
five-dimensional ${\cal N}=1$ theories via configurations of type IIB
$(p,q)$-fivebranes, stretched in a way such that 4+1 common dimensions of
the world-volume correspond to spacetime for the gauge theory.
In the transverse 5 directions the
branes look like lines which
conjoin in a web fashion.
We therefore suppress 3 of the transverse direction and
represent our branes as intersections of lines in a two-dimensional
co\"{o}rdinate system with axes $(p,q)$ so that the charge of the
$(p,q)$ fivebrane can be readily read out from its slope. 
To such configurations we shall refer as {\bf $(p,q)$-webs} of fivebranes.

The branes
intersect at a basic 3-valence vertex which we have drawn in Part
(a) of \fref{f:pqeg}. In general, the branes will extend in a skeletal
structure woven by 3-valence vertices at which $(p,q)$-charges are
conserved. That is, at each vertex, the sum of the $(p,q)$-vectors
vanishes. Furthermore, there are internal lines and external lines,
the former of which has no free ends; these are colour and flavour
branes respectively.
An archetypal example of the web is given in Part (b) of \fref{f:pqeg}.
In general, a web with $N_c$ parallel internal lines and
$N_f$ external lines corresponds
to an $SU(N_c)$ theory with $N_f$ flavours. 

\EPSFIGURE[ht]{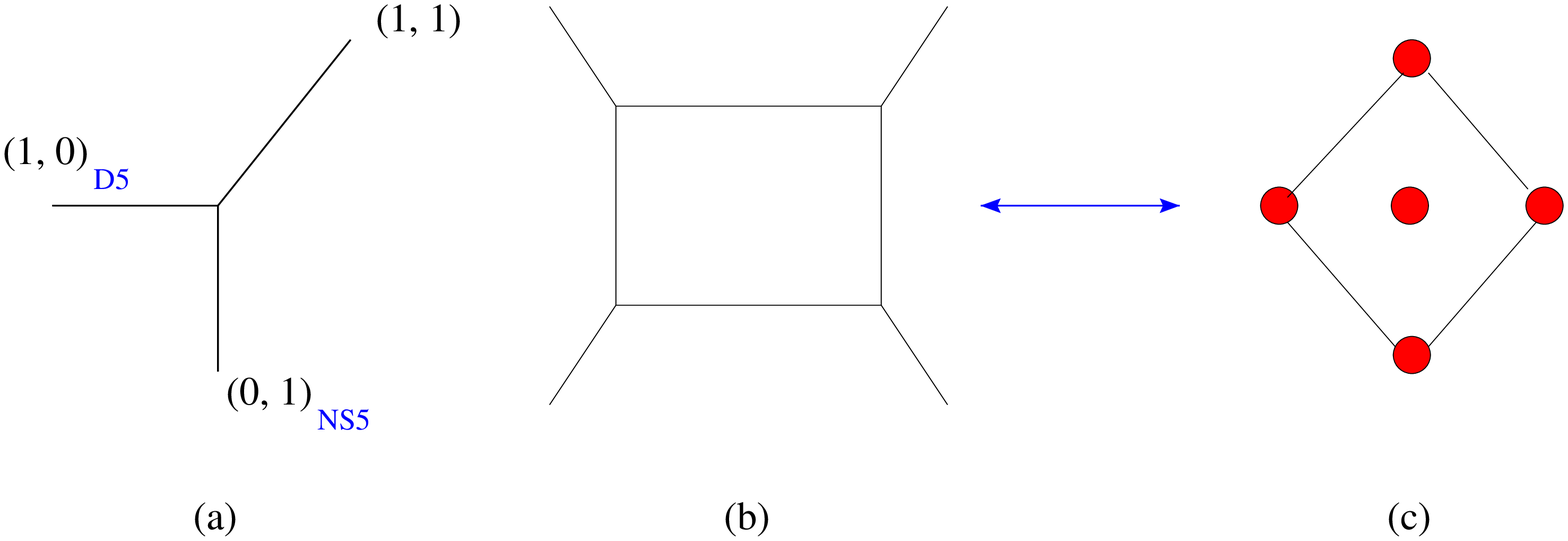,width=13cm}
{(a) The triple vertex of a D5-brane (1,0), an NS5-brane (0,1) and a
(1,1)-brane; (b) An archetypal example of a $(p,q)$-web and (c) its
dual grid (toric) diagram. 
\label{f:pqeg}
} 

The Higgs branch of the gauge theory is parametrised by the 
abovementioned three suppressed dimensions, 
while the Coulomb branch corresponds to
deformations of the relative positions of the $(p,q)$-branes. In (b)
of \fref{f:pqeg}, fundamental strings stretched between the parallel
fivebranes are BPS states: those between the horizontal correspond to
W-bosons of mass $T_s \Delta y$ and those between the vertical,
instantons of mass $|\tau| T_s \Delta x$ (where $\tau$ is the type IIB
scalar 
and $T_s$, the fundamental string tension).

The number of global deformations of  the theory is $SO(3)$ coming from
the Higgs branch; in addition there are
\[
n_G = \#(\mbox{external faces}) - 3
\]
extra global deformations.
On the other hand, the number of local deformations is
\[
n_L = \dim (\mbox{Coulomb Branch}) = \#(\mbox{internal faces}) \ .
\]
~\\
~\\

\subsection{Grid Diagrams, Toric Diagrams and Four-dimensional Gauge
Theory}
Having explained the fundamental rules of the representing
five-dimensional gauge theories with $(p,q)$-webs. We now explain how
this configuration gives rise to a four-dimensional ${\cal N}=1$ gauge
theory, which is in particular a world-volume theory of a D3-brane
probe of a toric singularity. For some rudiments of toric geometry in
the present context, the reader could read, for example,
\cite{thesis}. We point out that, of course, the toric diagrams we
draw should be three-dimensional because we are dealing with complex
threefolds. However, since we are dealing with local Calabi-Yau threefolds,
we henceforth draw only the
two-dimensional cross section of the toric diagram.
The endpoints of the vectors are
coplanar due to the Calabi-Yau condition.

As explained in \cite{Aharony:1997bh}, one could define
a dual diagram to the $(p,q)$-web, a so-called grid diagram. 
The duality is in the sense of finite planar graphs, where vertices
and faces are interchanged and edges, to their perpendiculars.
This procedure is illustrated in going from our prototypical example
(b) to its dual in (c) in \fref{f:pqeg}. The astute reader may
recognise (c) to be nothing but a toric diagram; this is precisely the
point of \cite{Aharony:1997bh,Leung:1997tw}. The Inverse Algorithm of
\cite{Feng:2000mi} was devised exactly for taking this data to gauge
theory data in terms of quiver diagrams.

Indeed, in the perspective of
\cite{Leung:1997tw}, it was noted that when geometrically engineering
gauge theories from toric varieties, a simple correspondence exists
between the geometry and branes. The key features of toric varieties
are captured by when the torus fibrations shrink. However, this is
precisely where the shrinking cycles become a source of the
brane-charges. Therefore these loci should be identified with
appropriate brane configurations. To be specific, we have identified
lines in the $(p,q)$-web with D5-branes. In addition, there could be
(fractional) D3 and D7 branes. Then, the D3-branes can be associated
with vertices and D7, with the faces. In other words, we let D3, D5
and D7 branes wrap 0, 2 and 4 cycles respectively in a Calabi-Yau
threefold geometry. The result, is a four-dimensional gauge theory,
which, on the D3-brane world volume perspective, is the  ${\cal N}=1$,
four-dimensional quiver gauge theory from probing the said
Calabi-Yau, which is a (non-compact) toric variety whose toric diagram
is specified by the grid diagram.

In light of these insights, obtaining the matter content of the
${\cal N}=1$ gauge theories on D-branes probing toric 
singularities can be extremely
intuitive and straight-forward.
Indeed, given the toric diagram, one simply has to identify it with a 
grid diagram, and then draw the dual; this is then the $(p,q)$-web.
Assigning appropriate $(p_i,q_i)$-charges to each external leg
$i$ according to the direction of the vectors,
the quiver matrix of the resulting theory is instantly given by
\beq
\label{interrule}
\chi_{ij} = \det \mat{p_i & p_j \cr q_i & q_j } \ . 
\eeq
Of course, there are
limitations to this methodology: 
in contrast to the generality of the Inverse
Algorithm of \cite{Feng:2000mi}, 
there is a present lack of a direct method to obtain the
superpotential. Moreover, subtleties arise when there are multiple internal
points (or equivalently, parallel external legs).
More pertinent to us is the fact that
we see that \eref{interrule} is naturally
antisymmetric in the indices $i$ and $j$. The
non-antisymmetric parts are (1) adjoint matter corresponding to
non-zero diagonal entries in the quiver matrix and (2) non-chiral
matter corresponding to bi-directional
arrows, i.e., symmetric parts, in the quiver matrix.
Capturing these fields, though being discussed nicely in
\cite{Hanany:2001py}, still lacks a
systematic analysis from geometric methods such as $(p,q)$-webs.

An important point is that the $(p,q)$-web
method culminating in \eref{interrule}
seems to suggest a one-to-one
correspondence between the toric data and the $(p,q)$-web data in
the sense that they are dual finite graphs. Once the toric diagram is
given, the $(p,q)$-web, and thence the quiver, are uniquely
determined. We will point below how one needs to be careful with this
identification and emphasise the concepts of ``toric'' versus
``quiver'' $(p,q)$-webs.

\newpage
%
\section{Case Studies: Partial Resolutions of $\IC^3/\IZ_3 \times
\IZ_3$}\label{s:case}
We will be focusing on the D-brane probe
theories that arise from partial resolutions of Abelian orbifolds,
especially the cones of the Hirzebruch and del Pezzo surfaces; these
have readily available data from the Inverse Algorithm with which we
may compare and contrast. Some of
these have been very nicely considered in
\cite{Hanany:2001py,Franco:2002ae} from the $(p,q)$ point of view.

The gauge theories on D-brane probes to partial resolutions of the
Abelian orbifold $\IC^3/\IZ_3 \times \IZ_3$ have been extensively
studied \cite{Chris1,Feng:2000mi}; the Inverse Algorithm of
\cite{Feng:2000mi} and subsequent duality considerations
\cite{Feng:2001xr,Feng:2001bn,Feng:2002kk,Feng:2002zw,Beasley:2001zp,unhiggs}
have
provided us with a wealth of explicit examples. The $(p,q)$-web
techniques reviewed in the previous Section have been applied to 
these examples in \cite{Franco:2002ae}.

In this section, we will re-examine these examples together with some
new ones, in order to
demonstrate that there are really two kinds of $(p,q)$-webs.
In particular we will focus on the cones over Hirzebruch surfaces and
so-called pseudo-del Pezzo surfaces.
\subsection{The $F_2$ Singularity}
We begin with an example not previously addressed in the
literature. This is the affine cone $F_2$ over the second Hirzebruch
surface. 
The surface itself is a $\IP^1$-fibration over $\IP^1$. 
$F_2$ is a local Calabi-Yau threefold which can be
described as a singular affine variety. The
toric diagram (with a perpendicular dimension of the cone suppressed)
for $F_2$ is given in part (b) of \fref{F2}. The
background configuration of blue dots is the familiar 
toric diagram of $\IC^3/\IZ_3 \times \IZ_3$. We see that there
is an embedding of toric diagrams. This means that the Inverse
Algorithm of \cite{Feng:2000mi} can be applied.

\EPSFIGURE[ht]{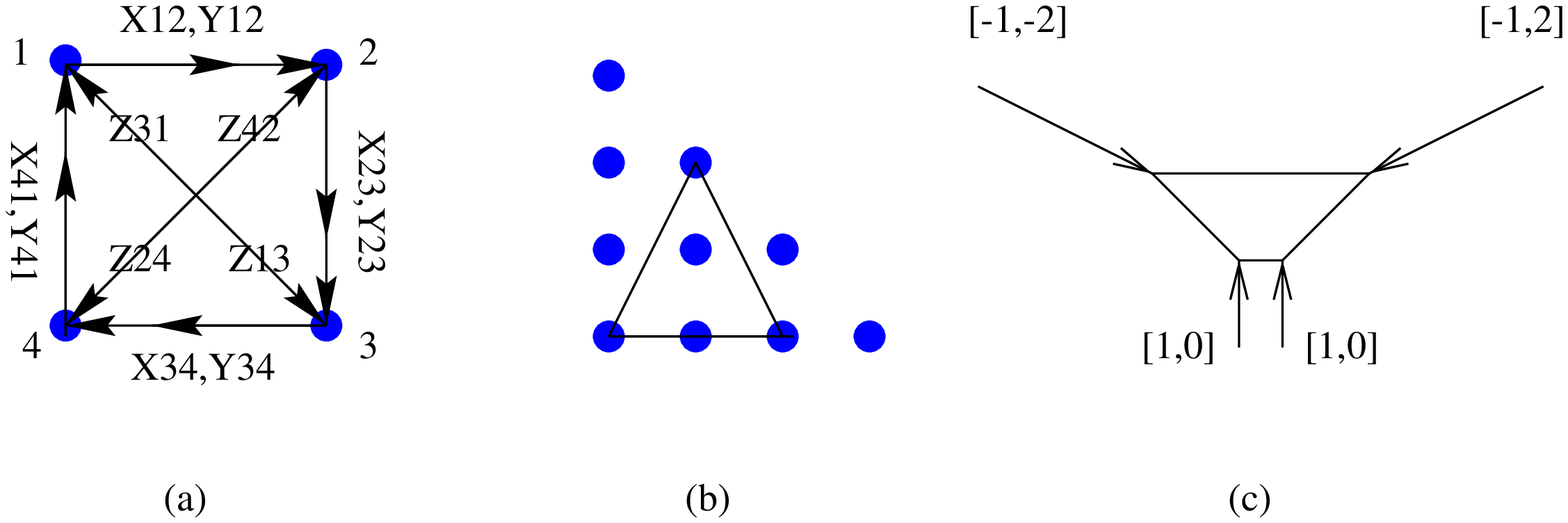,width=7in}
{The theory for the singularity $F_2$ which is a cone over the second
Hirzebruch surface: (a) the quiver diagram; (b) the toric diagram
(bounded by the lines) given as an
embedding into that of $\IC^3/\IZ_3 \times \IZ_3$;
(c) the corresponding $(p,q)$-web diagram.
\label{F2}
} 
Subsequently,
we can readily obtain the world volume theory of a brane probing $F_2$.
The matter content is given by the quiver diagram depicted in part (a)
of \fref{F2}; it is a theory with 4 product gauge groups and 12
bi-fundamental fields. The superpotential is given by
\bea
W_{F_2} & = & X_{12} Y_{23} Z_{31}-Y_{12} X_{23} Z_{31} + X_{23} Y_{34} Z_{41}
      -Y_{23} X_{34} Z_{41} \label{WF2}\\
 & + & X_{34} Y_{41} Z_{13}- Y_{34} X_{41} Z_{13} + X_{41} Y_{12} Z_{24}
      - Y_{41} X_{12} Z_{24} \ . \nonumber
\eea 

As discussed above, instead of using the Inverse Algorithm,
the matter content, at least, can be easily
obtained from a $(p,q)$-web configuration. We dualise the toric
diagram to obtain the web; the result is presented in 
part (c) of \fref{F2} with the $(p,q)$-charges appropriately labelled.
The quiver is then obtained by the intersection rule \eref{interrule}
from the $(p,q)$-charges. The answer is in perfect agreement with that
obtained from the Inverse Algorithm, presented in Part (a) of the
figure. 

Note that the procedure of obtaining the web from the toric diagram
involves a direct dualisation of the graph. We henceforth
call a web obtained this way
a {\bf toric $(p,q)$-web} to reflect the fact that it is directly and
uniquely obtained from
the toric diagram by graph-dualisation.

\begin{figure}[ht]
\centerline{
$A = \mat{0 & 2 & 1 & 0 \cr 0 & 0 & 2 & 1 \cr
1 & 0 & 0 & 2 \cr 2 & 1 & 0 & 0}, \qquad
\qquad \qquad
\ba{r}
  \epsfxsize = 6cm
  \epsfbox{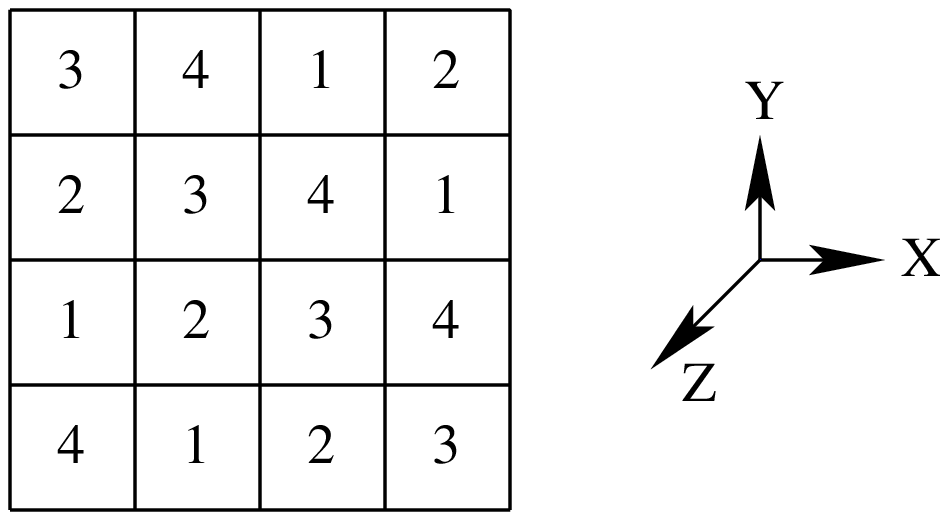}
\ea$}
  \caption{The Brane Box model of for the orbifold
theory $\IC^3/\IZ_{4}$ with action $(1,1,2)$. We have also included
the adjacency matrix $A$ for reference.}
  \label{F2box}
\end{figure}

Now, in presenting the quiver and the superpotential, we have 
purposefully chosen the notations above.
The reader may instantly recognise, upon observing the theory
presented in the said manner, that the theory is nothing
other than the familiar theory of the Abelian orbifold
$\IC^3/\IZ_{4}$ with action $(1,1,2)$. That is, with the action of the
$\IZ_4$ on the co\"{o}rdinates $(x,y,z)$ of $\IC^3$ as $(x,y,z)
\rightarrow (\omega_4 x, \omega_4 y, \omega_4^2 z)$, where $\omega_4$
is the primitive fourth root of unity. Quivers for Abelian orbifolds
are most conveniently given in terms of the Brane Box Model
representation \cite{BBM}. For comparison, we have drawn the 
Brane Box Model of this theory in \fref{F2box}.

Regarding this equivalence to the orbifold theory, we remark that
in general, the toric
diagram for $\IC^{3}/\IZ_{1+a+b}$ with action $(1,a,b)$ is given in
Figure \ref{Zab}.
\EPSFIGURE[ht]{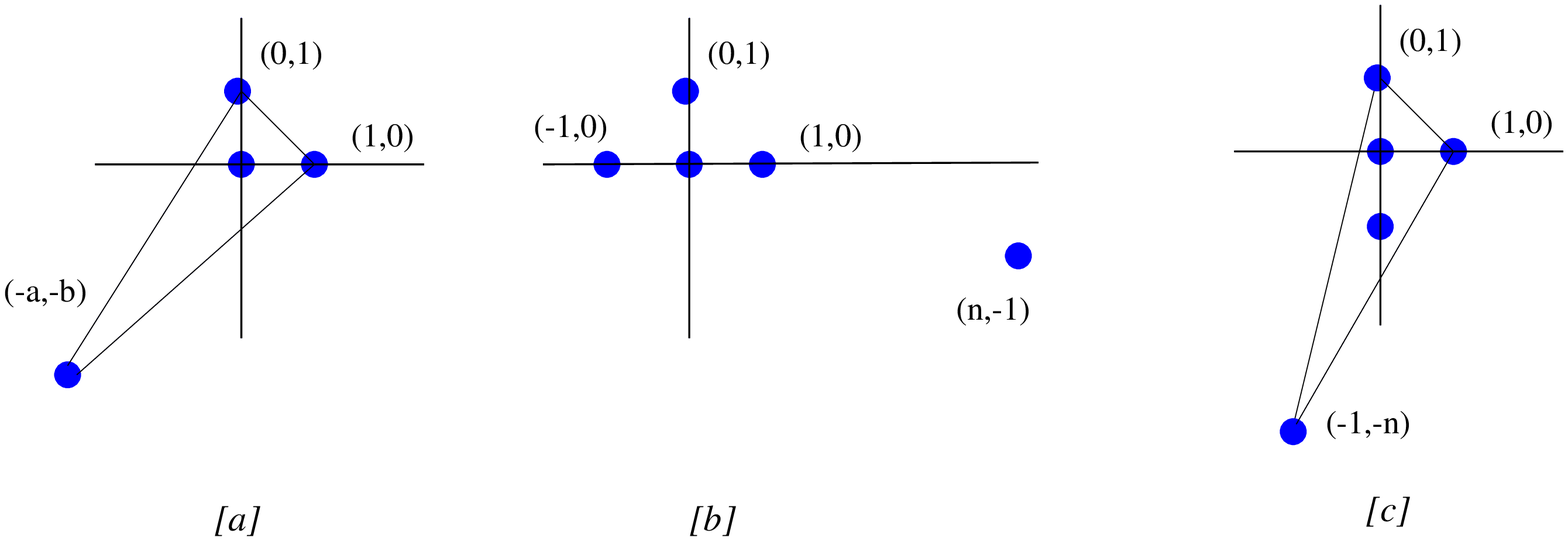,width=7in}
{
(a) The toric diagram of the weighted projective space
$\IP^2_{[1,a,b]}$, which is the exceptional divisor in the resolution
of the abelian orbifold
$\IC^3/\IZ_{1+a+b}$ with action $(1,a,b)$. 
(b) The toric diagram of the $n$-th Hirzebruch surface. (c) We redraw
(a) for the case of $\IP^2_{[1,1,n]}$ and see that toric diagram, up
to rotation, coincides with that of (b).
\label{Zab}
}
This toric diagram, when visualized in the two-dimensional cross
section, 
is that of the weighted projective space $\IP^2_{[1,a,b]}$. 
This fact was observed in \cite{Feng:2002kk},
and the corresponding ``toric'' $(p,q)$-web has also been
constructed therein. Indeed for large $(a,b)$, there may be multiple
interior points. This would therefore 
be an example which demonstrates that the one-to-one 
correspondence, in the sense of dual graphs, between the
toric $(p,q)$-web and the toric diagram is still valid even with
multiple inner points in the toric diagram.  The web diagram can
also be used to investigate the resolution of singularities, as 
discussed in \cite{Feng:2002kk}. 

More importantly, we have
that the quiver obtained from \eref{interrule}
for $F_{n \ge 2}$, the cone over the
$n$-th Hirzebruch surface,
is given simply by that of the Abelian orbifold 
$\IC^{3}/\IZ_{n+2}$ with action $(1,1,n)$.
We have used our toric algorithm to verify the cases of 
$F_3$ and $F_4$. This is both remarkable and unsurprising. It is
remarkable because we have reduced the matter content for a series of
complicated geometries to those of simple Abelian orbifolds:
\beq
\mbox{Quiver}(F_{n \ge 2}) = 
\mbox{Quiver}(\IC^{3}/\IZ_{n+2}), \quad
\mbox{with~action~}(1,1,n).
\eeq
It is also unsurprising because upon observing part (c) of \fref{Zab}
for the special case of $a=1, b=n$, we see that for $n \ge 2$, 
the toric diagrams for $F_n$ and $\IP^2_{[1,1,n]}$
are the same (up to rotation). 
Indeed, as $\IP^2$ (which, incidentally, is the zeroth del Pezzo
surface) is the exceptional divisor in the resolution of
$\IC^{3}/\IZ_3$, so too is the weighted projective space
$\IP^2_{[1,1,n]}$ the exceptional divisor for the resolution of
$\IC^{3}/\IZ_{n+2}$.  Now, because we have shown that the $(p,q)$-webs
of $F_n$ and $\IP^2_{[1,1,n]}$ are identical for $n \ge 2$, it is not
surprising that the quiver for $F_n$, or at least the antisymmetric part thereof
obtained from \eref{interrule}, coincides with that of 
$\IC^{3}/\IZ_{n+2}$. As a parenthetical note, construction of
exceptional collections of coherent sheafs over $\IP^2$ has been
central to D-brane interpretations of the McKay correspondence
\cite{McKay,McKay2}, generalisations to the aforementioned
weighted projective spaces would be interesting.

We mentioned earlier that the $(p,q)$-web technique readily gives the
antisymmetric part of the quiver. We
note that there are two bi-directional arrows in the quiver
diagram (a) of \fref{F2} obtained from the Inverse Algorithm. 
If we neglect these bi-directional arrows, 
the diagram is something we have seen
before: it is exactly phase II of theory probing $F_0$, the cone over
the zeroth Hirzebruch surface! The reader is referred to (2.2) of
\cite{Feng:2001xr} or Figure 4 of \cite{Feng:2002zw} for this
theory. For completeness let us include the relevant data here for the
readers' convenience. The figure contains the respective quivers of and
the superpotentials are tabulated for $dP1$, the cone over the first
del Pezzo surface, the two Seiberg dual phases $dP2_I$ and $dP2_II$
for $dP2$, the cone over the second del Pezzo surface, as well as the
two phases $F0_{I}$ and $F0_{II}$ of $F_0$, the cone over the zeroth
Hirzebruch surface.
%
\bea
&&\hspace{-1in}\epsfxsize = 15cm\epsfbox{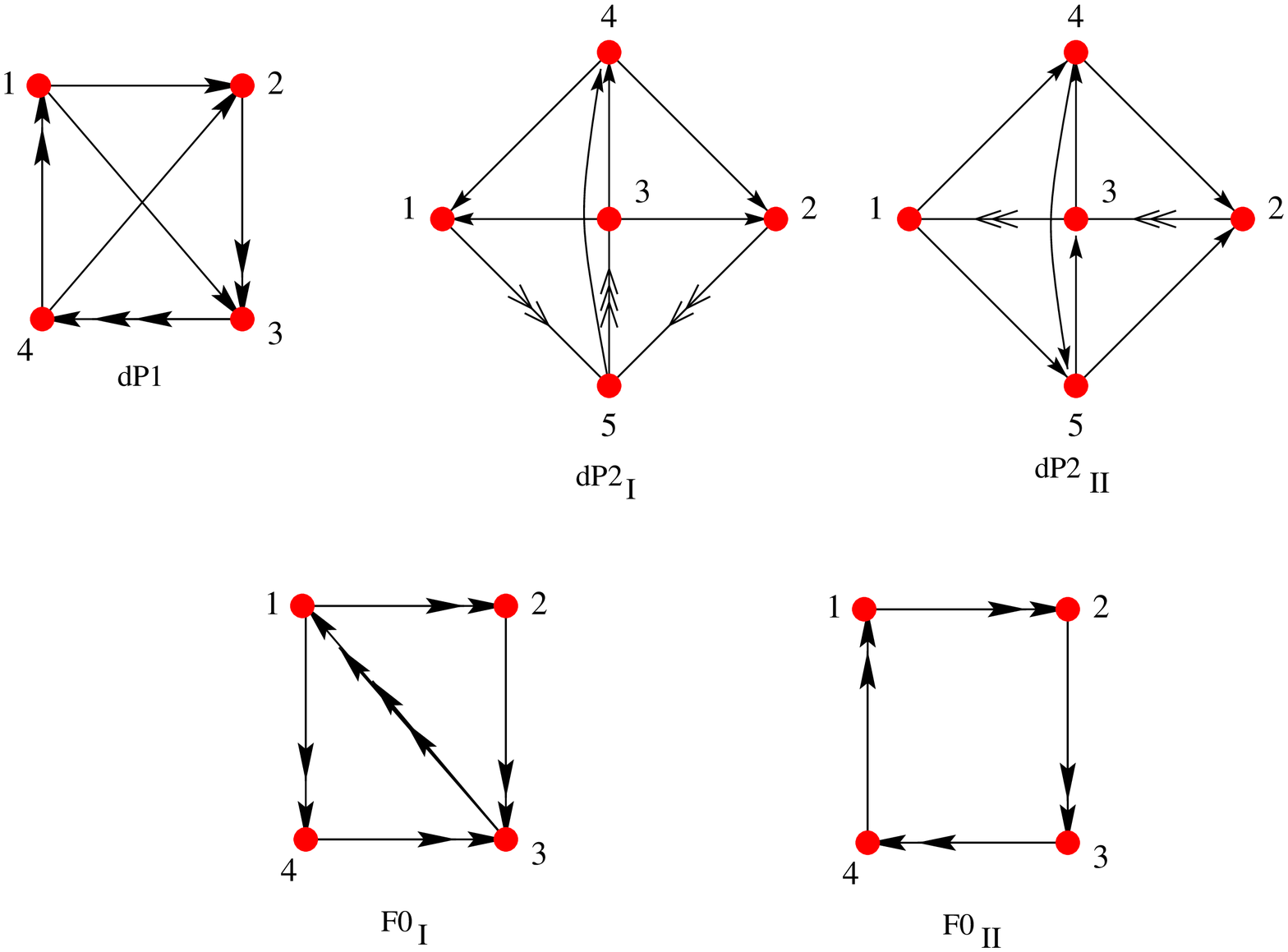}\nonumber \\
W_{dP1} & = & X_{34} Y_{41} X_{13}- Y_{34} X_{41} X_{13} -X_{34}
X_{42} Y_{23} + Y_{34} X_{42} X_{23} \nonumber \\
&& \quad - X_{12} X_{23} Z_{34} Y_{41} + X_{12} Y_{23} Z_{34} X_{45},
\nn \\
W_{F0_I} & = & X_{14}^1 X_{43}^1 M_{31}^{22}- X_{14}^2 X_{43}^1 M_{31}^{12}
-X_{14}^1 X_{43}^2 M_{31}^{21}+X_{14}^2 X_{43}^2 M_{31}^{11}\nonumber \\
&& \quad - X_{12}^1 X_{23}^1 M_{31}^{22}+X_{12}^2 X_{23}^1 M_{31}^{12}+
X_{12}^1 X_{23}^2 M_{31}^{21}-X_{12}^2 X_{23}^2 M_{31}^{11}, \nn \\
W_{F0_{II}} & = & X_{12} X_{23} Y_{34} Y_{41} -Y_{12} X_{23} X_{34} Y_{41} 
-X_{12} Y_{23} Y_{34} X_{41} +Y_{12} Y_{23} X_{34} X_{41},\nn \\
W_{dP2_I} & = & [X_{41} X_{15} X_{54}-X_{42} X_{25} X_{54}]
 -[X_{41} Y_{15} X_{53} X_{34}-X_{42} Y_{25} Y_{53} X_{34}]\nonumber \\
& & \quad - [X_{31} X_{15} Y_{53}-X_{32} X_{25} X_{53}]
+[ X_{31} Y_{15} Z_{53}-X_{32} Y_{25} Z_{53}],\nn \\
W_{dP2_{II}} & = & [X_{34} X_{45} X_{53}] - [X_{53} Y_{31} X_{15} 
+X_{34} X_{42}Y_{23}]\nonumber\\ 
& &   +[Y_{23} X_{31} X_{15} X_{52}+
X_{42} X_{23} Y_{31} X_{14}] -[X_{23} X_{31} X_{14} X_{45} X_{52}].
\label{summary}
\eea

This above observation on the bi-directional arrows will be our
initiation into the concept of ``quiver
$(p,q)$-webs.''
We first define, given a quiver diagram, the notion of the
{\bf quiver $(p,q)$-web}; this is simply
the $(p,q)$-web diagram which produces a given quiver by rule
\eref{interrule} regardless of geometry. We emphasise again that 
since \eref{interrule} is an antisymmetric form,
it is exactly these bi-directional arrows that {\sl can
not be captured by the $(p,q)$ technique}. Therefore, as far as
$F_0$ and $F_2$ are concerned, they have the same ``quiver
$(p,q)$-web'' even though their ``toric $(p,q)$-webs'' differ because
their toric diagrams obvious differ and we recall that the toric
$(p,q)$-web is obtained from the toric diagram by graph dualisation.
In the geometry this means
that the intersection numbers among vanishing cycles
(cf.~\cite{Feng:2002kk}) are antisymmetric for these theories and
cannot capture non-chiral matter. 
We now exploit this ambiguity in the correspondence
between the quiver $(p,q)$-web diagram and the toric data in detail.

\subsection{The $PdP2$ Singularity}
Our next example is what was called in Section 6 of \cite{unhiggs} as
the cone over a ``Pseudo del Pezzo surface.'' Recall, we call a surface
pseudo del Pezzo if it is toric and obtained by a single blowup
(inclusion of a point in the toric diagram) of a toric
del Pezzo surface,  but is not itself a del Pezzo surface 
because the blowup point is not
in general position. In other words,
it is $\IP^2$ blown up at non-generic points.
Indeed, though only the first 3 del Pezzo surfaces
are toric varieties, that is, $\IP^2$ blown up at up to 3 generic
points are toric, a higher number of blowups can be included and the
surface remains toric so long as these blowups are in special
positions.

We illustrate the first case of a cone over
a Pseudo del Pezzo surface, namely
$PdP2$, in \fref{dP1todP2}. In part (a), we have the toric
diagram of our
familiar $dP1$, the cone over the first del Pezzo surface
(cf.~e.g.~\cite{Feng:2000mi} and \cite{thesis}), embedded into 
$\IC^3/\IZ_3\times \IZ_3$. 
Under the condition that there is only one interior point, we can
blow up $dP1$ in two ways, as shown in (b) and (c). Part (c) is the
familiar $dP2$, the cone over the second del Pezzo surface while (d)
is what we call $PdP2$, the cone over the second pseudo del Pezzo
surface.
\EPSFIGURE[ht]{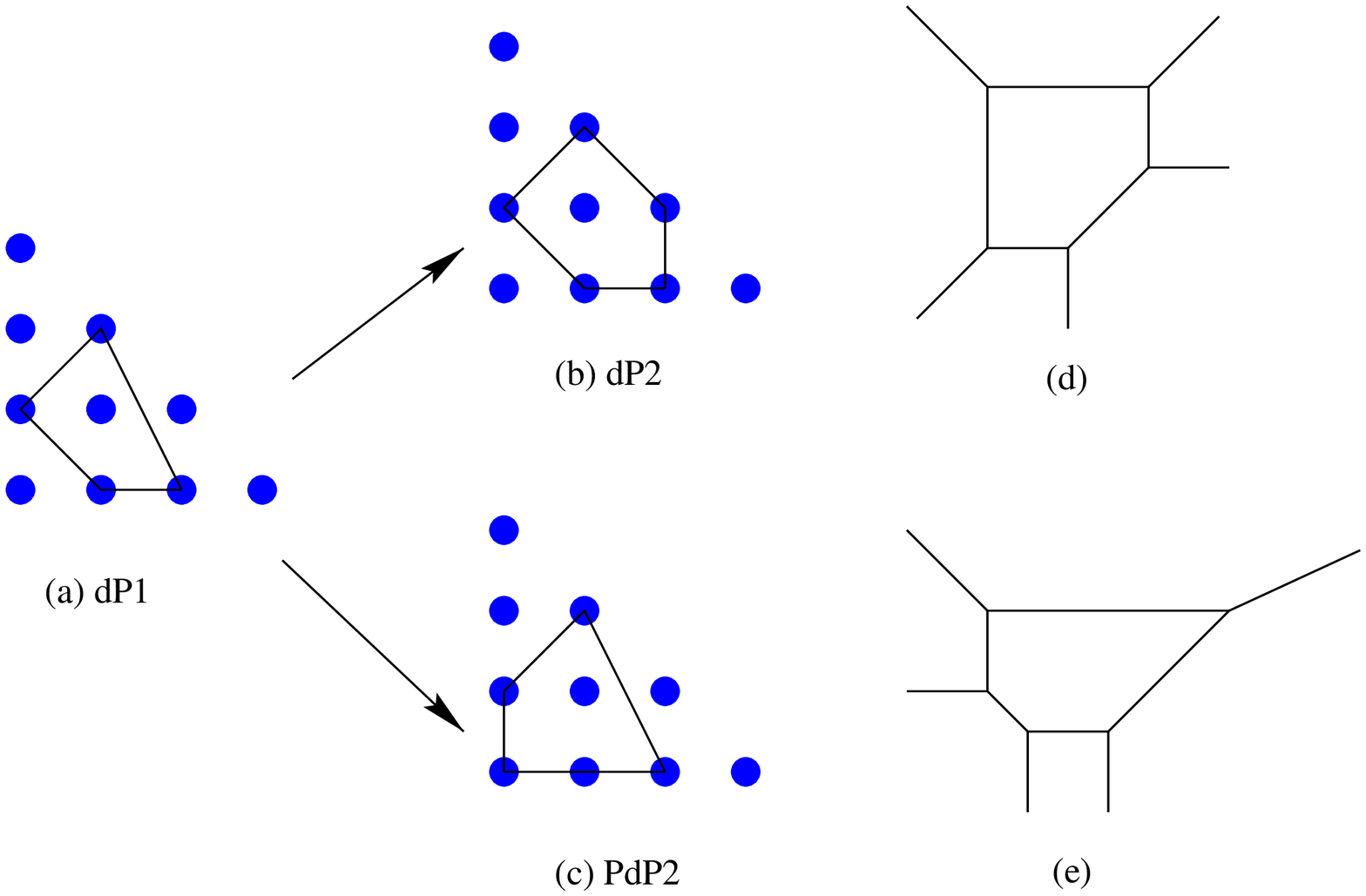,width=15cm}
{(a) The toric diagram of $dP1$, embedded into that of
$\IC^3/\IZ_3\times \IZ_3$. $dP1$ can be blown up in two ways (in the
toric diagram this is done by the
inclusion of another point) into (b) $dP2$, the cone over the second
del Pezzo surface, or (c) $PdP2$, the cone over the second pseudo del
Pezzo surface. The toric $(p,q)$-web diagrams are drawn respectively
in (d) and (e).
\label{dP1todP2}
} 
The corresponding $(p,q)$-web diagrams are given respectively
in parts (d) and (e) of Figure \ref{PdP2}.  We remind the reader that
these, in our nomenclature, are toric
$(p,q)$-webs because they are obtained directly from dualising the
toric diagrams. Indeed (d) and (e) differ because the respective toric
diagrams differ; this is illustrative of the fact that
toric data and toric $(p,q)$-web diagrams should be in
one-to-one correspondence.

Now, in \cite{Franco:2002ae}, these two $(p,q)$-web
were obtained by the technique of splitting a brane. 
There, these two web diagrams were identified
as those for the two different toric dual phases of $dP2$.
We now point out that the web diagrams in fact correspond to different
geometries and that it is important to distinguish toric and quiver
$(p,q)$-webs. With some foresight, we note that
the condition of \cite{Franco:2002ae} 
for splitting one external
leg, such that the new external legs should not intersect with nearby
external legs, is exactly the condition for addition of points in the
toric diagram. 
We will return to this issue in a later section on splitting
webs.  We shall see that higgsing in the field theory, i.e., the
combining of nearby external legs, is exactly the reverse process of
deleting some nodes from the toric diagram to reach another
\cite{unhiggs}.
In other words, the splitting and combining of external legs in the
$(p,q)$-web language correspond naturally to embedding and deleting
nodes in toric diagrams, and thus, to blowing up and down in the
geometry.  
From this geometric point of view, it is natural that
there is a one-to-one 
correspondence between the (toric) $(p,q)$-web and toric diagram.
%
\EPSFIGURE[ht]{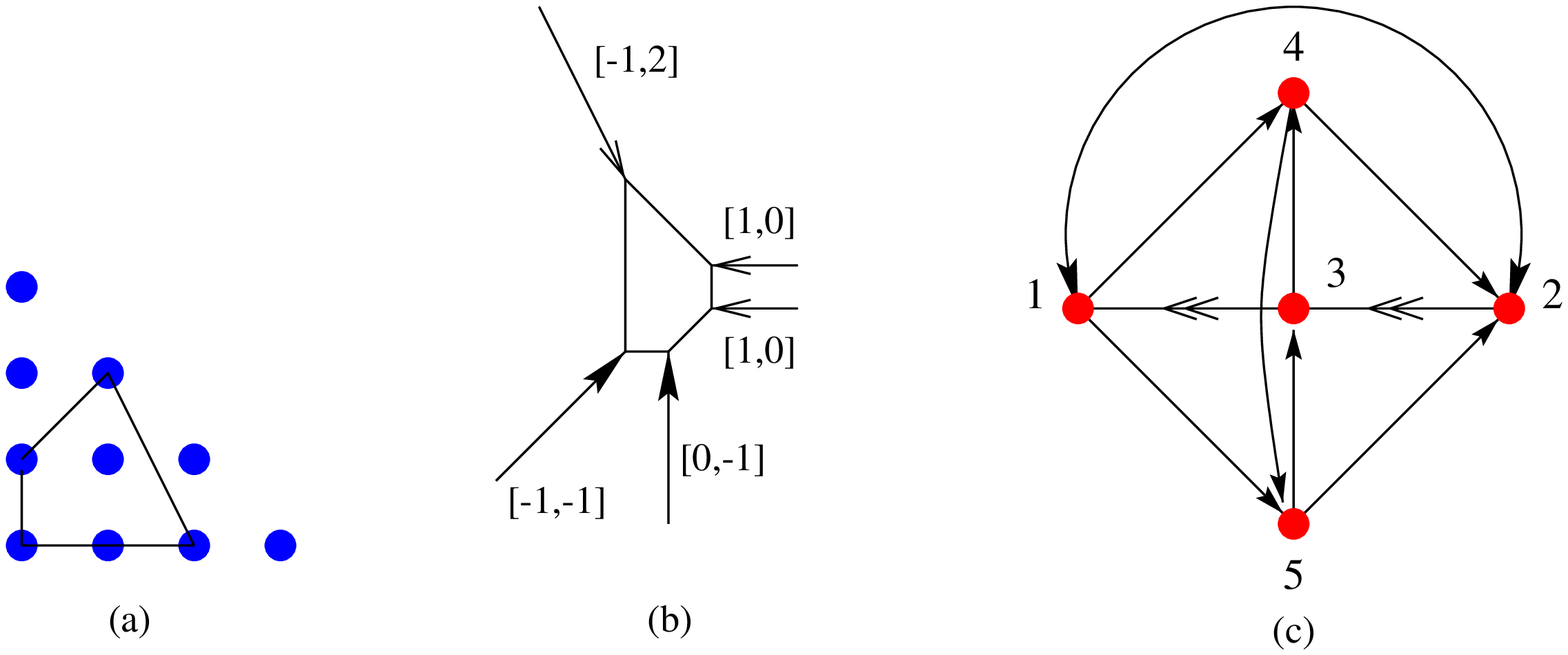,width=15cm}
{(a) The toric data of $PdP2$ and (b) its (toric) $(p,q)$-web diagram;
these are redrawn from \fref{dP1todP2}.
(c) the corresponding quiver diagram.
\label{PdP2}
} 

Now, let us return to the issue on bi-directional arrows and the
comparison between the two webs (d) and (e) in \fref{dP1todP2}.
The quiver diagram of $PdP2$ singularity is given in part (c) 
of Figure \ref{PdP2}. Notice that this quiver diagram is exactly
like phase II of $dP2$ given in the diagram in \eref{summary}, if
we ignored the bi-directional arrow between nodes 1 and 2. 
Once again, as in the case of $F_2$ and $F_0$ mentioned in the
previous subsection, ignoring bidirectional arrows identifies two
different theories.
Indeed, if we calculated the quiver diagram directly from the corresponding
$(p,q)$-web diagram in part (b) of \fref{PdP2}, 
we would have obtained phase II of $dP2$ and 
this is the reason why \cite{Franco:2002ae} identified this web as
that for $dP2_{II}$. Of course, this is a vestigial feature of the
fact that \eref{interrule} does not capture bi-directional arrows. 
We therefore need to be careful that the ``toric'' $(p,q)$-web
for $PdP2$ given in Part
(b) of \fref{PdP2} does not produce the right quiver for the theory
but instead gives that of $dP2_{II}$.

We can further see the difference by noting that
the superpotential for $PdP2$, from the Inverse Algorithm, 
given in terms of its 13 fields is
\bea
W_{PdP2} & = & X_{31} X_{15} X_{53}- X_{23} X_{34} X_{42} + Y_{31} X_{12}
  X_{23}- Y_{23} X_{31} X_{12} \nn \\
  & + & X_{14} X_{42} X_{21} -X_{15} X_{52} X_{21} -Y_{31}X_{14} X_{45} X_{53}
+ Y_{23} X_{34} X_{45} X_{52} \ , \label{WPdP2}  
\eea
which of course differs from $W_{dP2_{II}}$.
Incidentally, we can see that this theory is invariant under the action:
$1\leftrightarrow 2,4\leftrightarrow 5$ plus conjugation (i.e.,
reversal of arrows) of all fields
(cf.~\cite{Feng:2002zw} for discussion on these symmetries).
Moreover, we can find all the toric dual phases for $PdP2$.
To remain rank 1 after Seiberg duality (cf.~\cite{Feng:2001xr} 
for this condition), only nodes 4 and 5 are
suitable. Choosing either one (they are equivalent to each other by
symmetry), we get back to the same theory. 
Hence, there is only one toric dual phase for $PdP2$, as opposed to
$dP2$, which has two toric dual phases, as shown in \eref{summary}

\subsection{The $PdP3$ Family of Singularities}
Having seen the examples of $F_0$ versus $F_2$ and $dP2_{II}$ versus
$PdP2$, let us move on.
Now, starting from the toric diagram of $dP2$, we can embed it into three
different toric diagrams by adding one more node. This is given 
in \fref{dP2todP3}. 
The first is our familiar cone over the third del Pezzo surface $dP3$
(cf.~Figure 4 of \cite{Feng:2000mi}). The other two, in the convention
above, are also what we call pseudo del Pezzo's; these we respectively call
$PdP3b$ and $PdP3c$. We have also drawn their corresponding (toric)
$(p,q)$-web diagrams. We call these three members the $PdP3$ family
and will address them individually.
Later, as a check on the embedding of the toric data, 
we will demonstrate that higgsing any of $dP3$, $PdP3b$ and
$PdP3c$ in the spirit of \cite{unhiggs} gives back
the $dP2$ field theory.

\EPSFIGURE[ht]{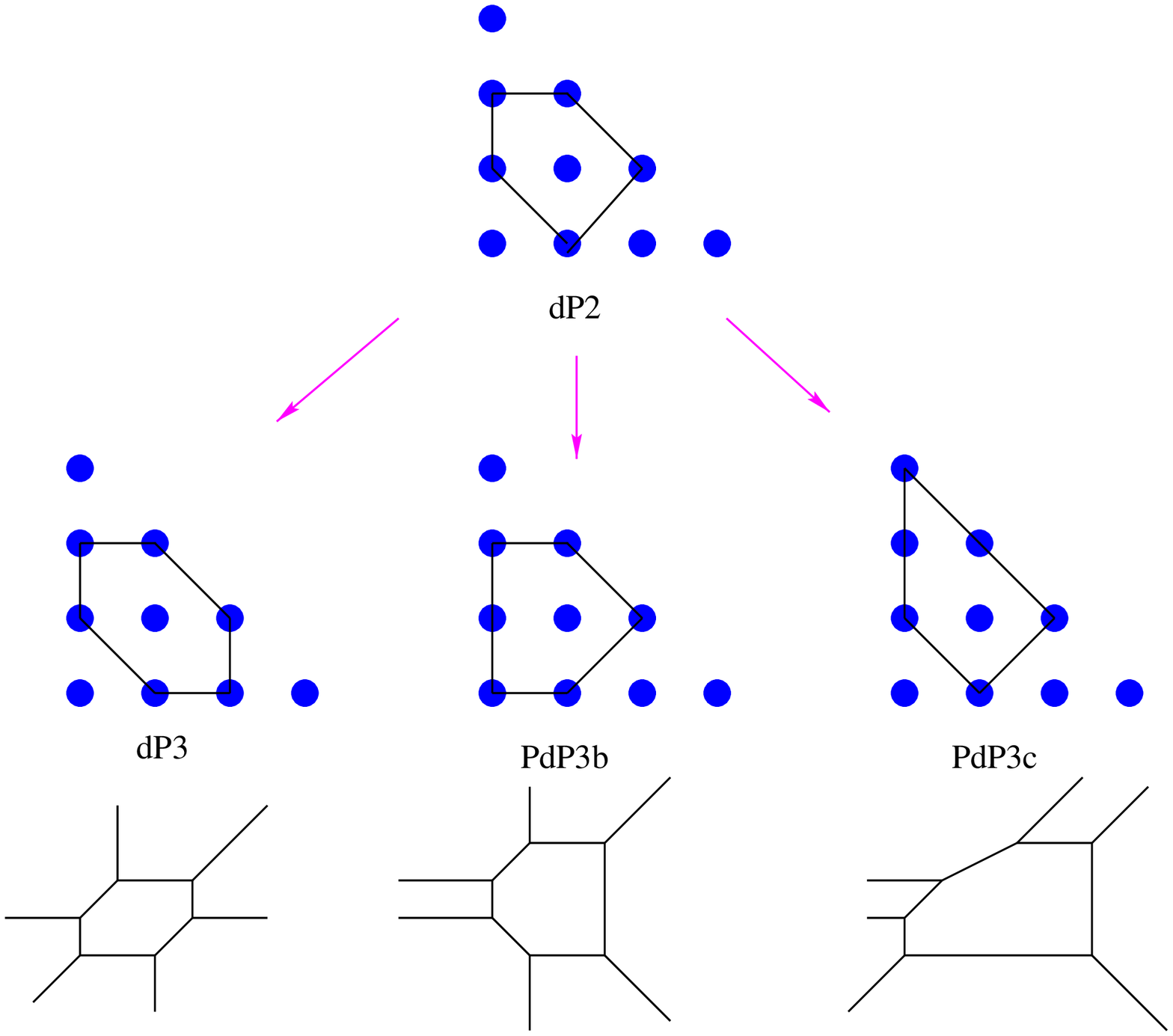,width=14cm}
{The possible three embeddings (unhiggsings)
of $dP2$ into $dP3$, $PdP3b$ and
$PdP3c$. The corresponding (toric) $(p,q)$-webs are also drawn.
\label{dP2todP3}
} 

\subsubsection{The $PdP3b$ Singularity}
Let us first analyse $PdP3b$.
The Inverse Algorithm applied to $PdP3b$ from partial resolution of
$\IC^3/\IZ_3\times \IZ_3$ results in three toric theories that
are Seiberg (toric) dual to each other. 
The quiver diagrams and the superpotentials for these three dual
phases are given below:
\bea
&&\epsfxsize = 14cm \epsfbox{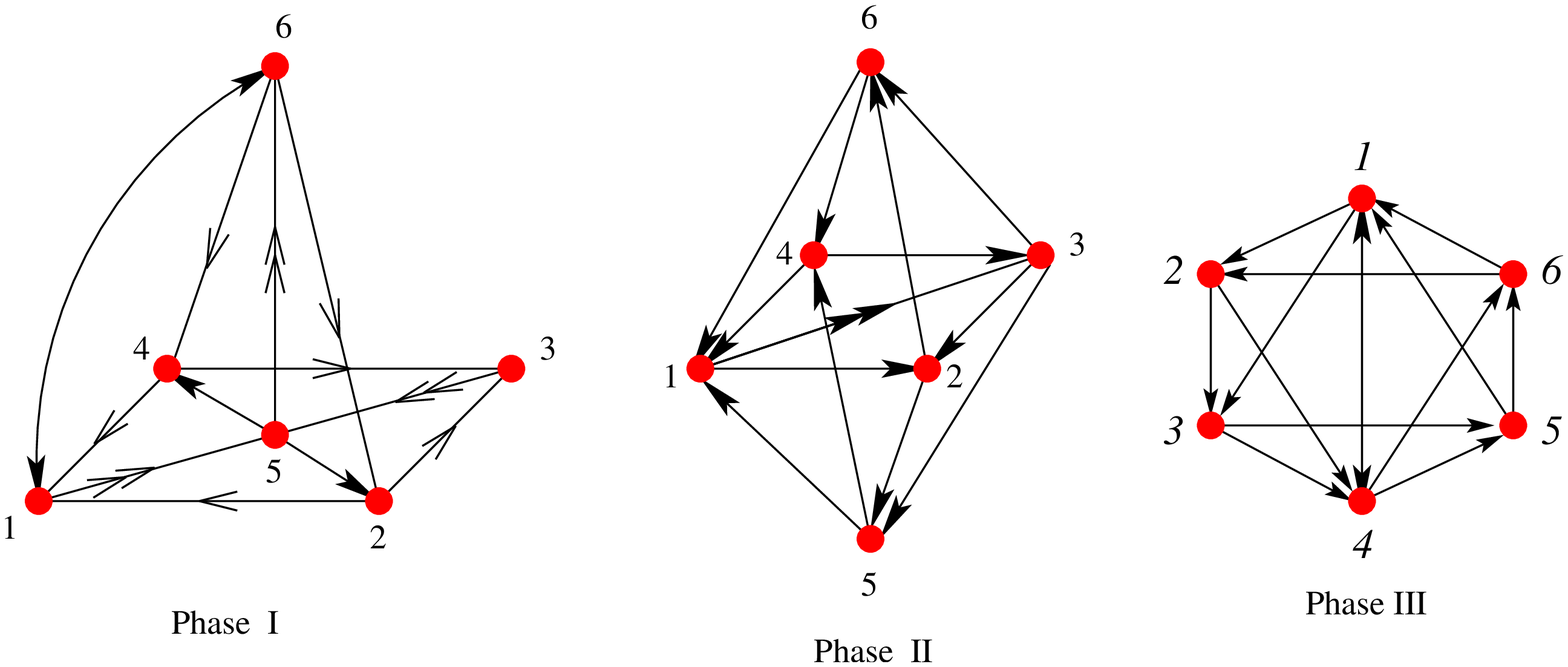}\nn\\
W_{PdP3b_I} & = &  X_{61} X_{15} Y_{56} - X_{61} Y_{15}  X_{56}
+ X_{16} X_{64} X_{41} -X_{16} X_{62} X_{21} \nn \\
& & \quad -X_{41} X_{15} X_{54} + X_{21} Y_{15} X_{52} +Y_{35} X_{54} X_{43} 
-Y_{35} X_{52} X_{23} \nonumber \\
&& \quad + X_{56} X_{62} X_{23} X_{35} - Y_{56} X_{64} X_{43} X_{35},
\nonumber \\
W_{PdP3b_{II}} & = & X_{12} X_{25} X_{54} X_{41} + X_{26} X_{64} X_{43} X_{32}
- X_{25} X_{51} Y_{13} X_{32} -  X_{64} X_{41} X_{13} X_{36} 
\nn \\
& & \quad + Y_{13} X_{36} X_{61} + X_{13} X_{35} X_{51} - X_{61} X_{12} X_{26}
-X_{43} X_{35} X_{54}, \nonumber \\
W_{PdP3b_{III}} & = & X_{13} X_{34} X_{41} -X_{46} X_{61} X_{14} + X_{45}
X_{51} X_{14} - X_{24} X_{41} X_{12}
+ X_{62} X_{24} X_{46} \nn\\
&& \quad -X_{35} X_{51} X_{13} + X_{23} X_{35} X_{56} 
X_{61} X_{12} - X_{23} X_{34} X_{45} X_{56} X_{62} \ .
\label{PdP3b}
\eea

We first remark on the bi-directional arrows
in comparison with the 4 phases of
our familiar $dP3$ theory. We will not present those quivers here
due to their similarity to the ones above and
the reader is referred to Figure 9 of \cite{Feng:2002zw}.
In summary,
\beq
\ba{|c|c|}\hline
PdP3b 	& dP3 \\ \hline
II	& II  \\ \hline
I	& III  \\ \hline
III	& I  \\ \hline
\ea
\eeq
where we have juxtaposed the quivers of the two theories that are
identical if we ignored the bi-directional arrows.
We remark also that
because of the bi-directional arrow in $PdP3b_I$, the quiver 
has only a $\IZ_2$ symmetry with action $2\leftrightarrow 4$ and we write
the superpotential in a manner to make the symmetry manifest.
The same holds for $PdP3b_{III}$; the
superpotential preserves only a $\IZ_2\times \IZ_2$ 
subgroup of the full $D_6$ symmetry that is expect of the quiver for 
$dP3$. Indeed, recalling the symmetry argument in  
\cite{Feng:2002zw}, it was shown that the $\IZ_2\times \IZ_2$ symmetry
uniquely determines the superpotential of phase II of $dP3$. 
Now since $PdP3b_{II}$ has the same quiver diagram, 
we should have a different symmetry in order to distinguish it. 
It is easy to check that this is indeed true and the phase II has
only a diagonal $\IZ_2$ symmetry with action $5\leftrightarrow 6,
1\leftrightarrow 3,2\leftrightarrow 4$ plus charge conjugation.
We conclude that $PdP3b$ and $dP3$ are indeed markedly different
theories even though they share the same quiver $(p,q)$-webs.

\subsubsection{The $PdP3c$ Singularity}
Our next member of the family is $PdP3c$, onto which we now move.
The toric diagram and corresponding $(p,q)$-web diagram were
given in Figure \ref{dP2todP3}.
In this case, we have two different phases, as obtained from the
Inverse Algorithm, with quivers drawn in \fref{PdP3c} and 
superpotentials given
as follows:
\bea
&&\epsfxsize = 11cm \epsfbox{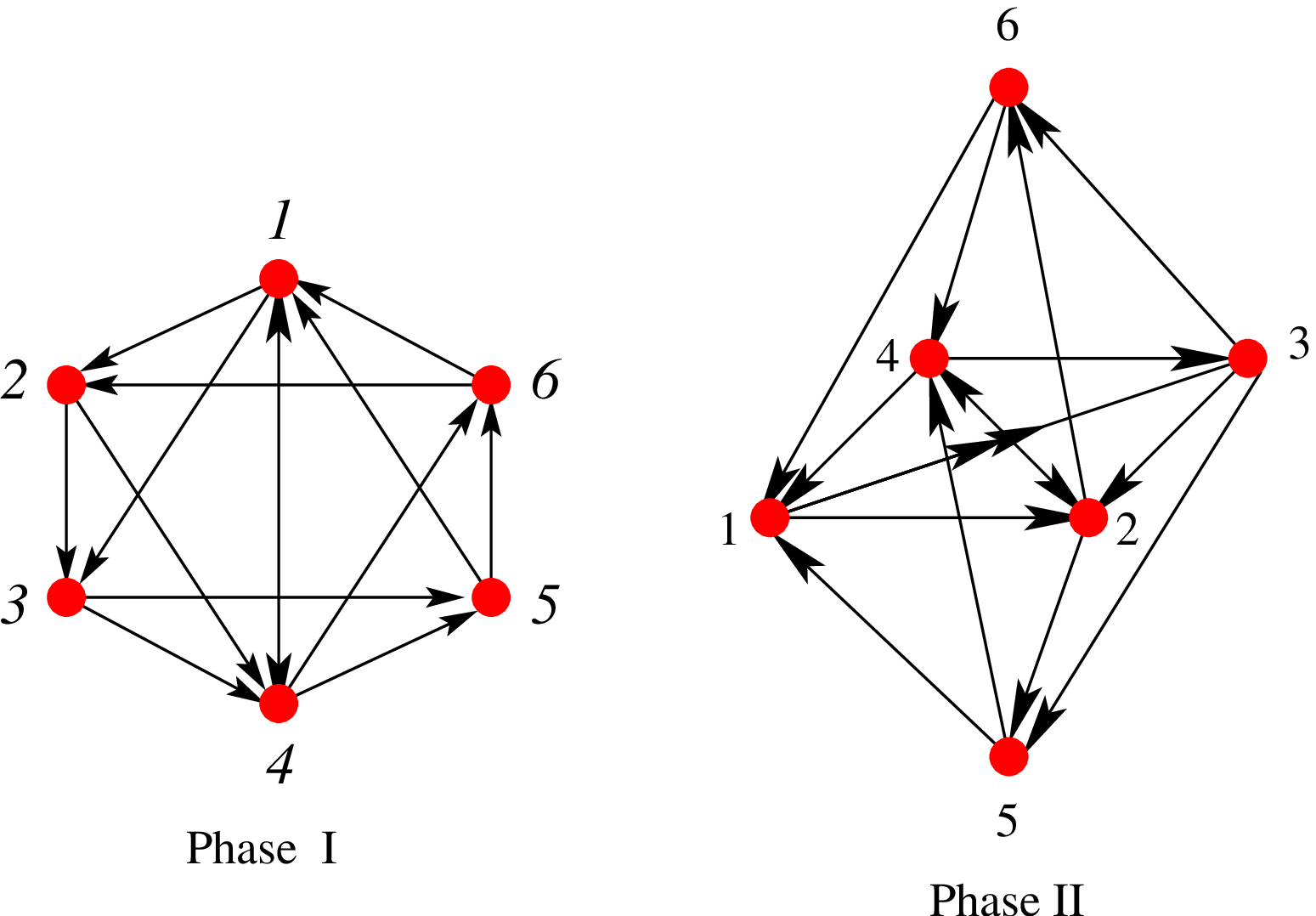}\nn\\
W_{PdP3c_I} & = & X_{12} X_{24} X_{41} + X_{45} X_{51} X_{14}-X_{13} X_{34}
	X_{41} - X_{46} X_{61} X_{14} \nn \\
&& \quad + X_{13} X_{35} X_{56} X_{61} + X_{46} X_{62} X_{23} X_{34} - X_{12} 
X_{23} X_{35} X_{51} - X_{45} X_{56} X_{62} X_{24}, \nonumber \\ 
W_{PdP3c_{II}} & = & X_{64} X_{42} X_{26} - X_{54} X_{42} X_{25} +X_{54} X_{43}
 X_{35} -X_{26} X_{61} X_{12} \nn \\
& & \quad + X_{24} X_{41} X_{12} - X_{24} X_{43} X_{32} + 
	X_{13} X_{36} X_{61} - X_{13} X_{35} X_{51} \nonumber \\
& & \quad + X_{25} X_{51} Y_{13} X_{32} - X_{41} Y_{13} X_{36} X_{64}
\ .
\label{PdP3c}
\eea
Again, let us compare with the phases of $dP3$.
For $PdP3c_I$, its quiver diagram is exactly the same as the phase I
of $dP3$ except for the bi-directional arrow between nodes 1 and 4. 
This arrow breaks the $D_6$ symmetry of the original $dP3$ quiver
down to $\IZ_2\times \IZ_2$, where one $\IZ_2$ takes $1\leftrightarrow
4, 2\leftrightarrow 5,3\leftrightarrow 6$ and the other takes
$2\leftrightarrow 6,3\leftrightarrow 5$ plus charge conjugation for
all fields.
It is easy to see these remnant symmetries from the
superpotential $W_{PdP3c_I}$ given in (\ref{PdP3c}).
For $PdP3c_{II}$, the quiver diagram is the
same as the phase II of $dP3$ except for the 
bi-directional arrow between nodes 2 and 4. This theory has only 
a diagonal $\IZ_2$ symmetry with action $1\leftrightarrow 3,
2\leftrightarrow 4,5\leftrightarrow 6$ plus conjugation.

\subsubsection{The $PdP3a$ Singularity}
The reader may question why we have not named a $PdP3a$
singularity. We now turn precisely to this issue.
The above theories of the $PdP3$ family were obtained from 
blowing up $dP2$; now let us
examine what happens when we blow up $PdP2$ from the toric diagram in
Part (a) of \fref{PdP2}. We present 
all possible blowups of the $PdP2$ toric diagram that still embed into
$\IC^3/\IZ_3\times \IZ_3$ in Figure \ref{PdP2toPdP3}.

\EPSFIGURE[ht]{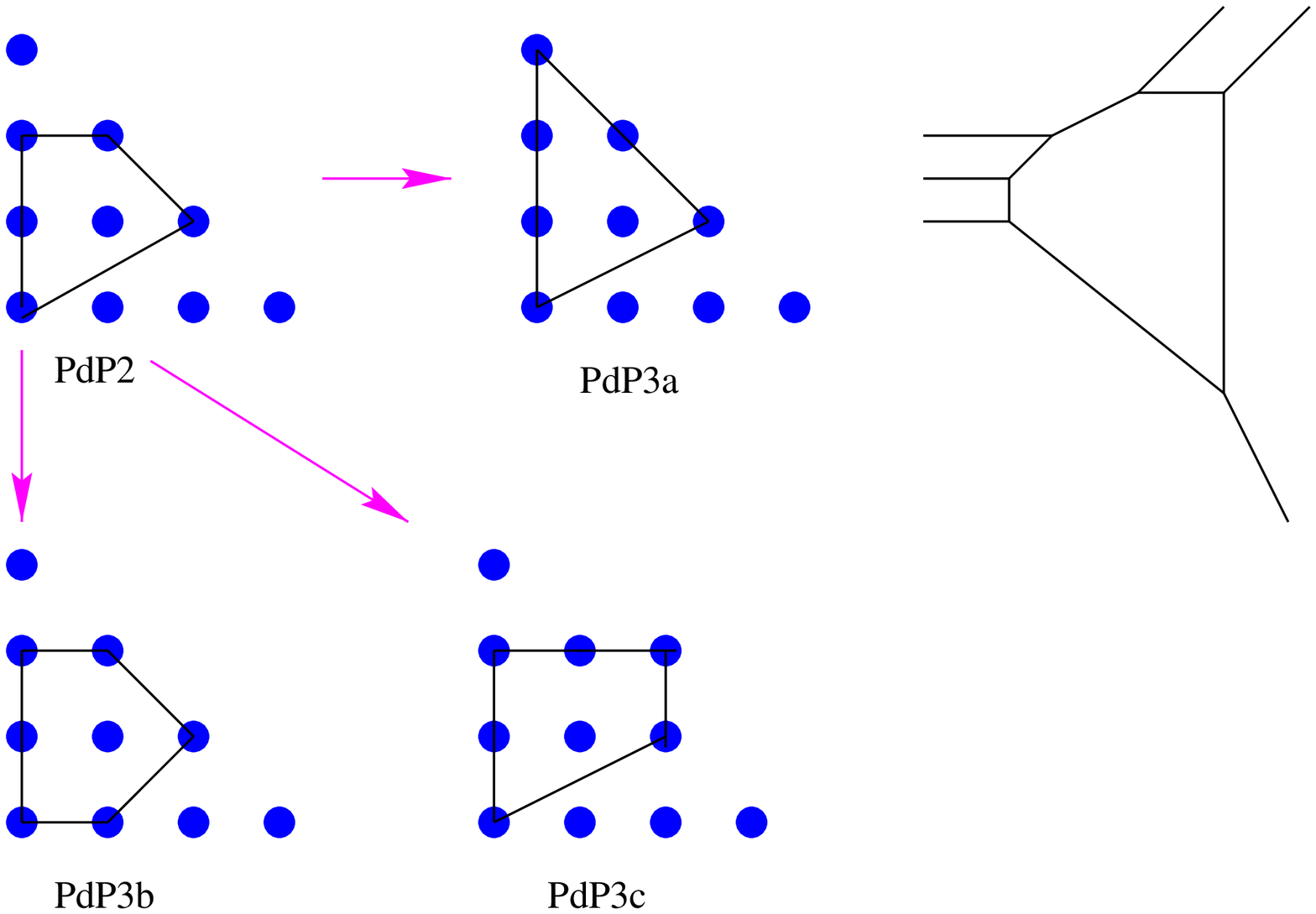,width=12cm}
{The three blowups of $PdP2$ into $PdP3a$, $PdP3b$ and $PdP3c$, which
still embed into the parent $\IC^3/\IZ_3\times \IZ_3$. The toric web
diagram for $PdP3a$ is also drawn. The toric $(p,q)$ webs for $PdP3b$
and $PdP3c$ are given in \fref{dP2todP3}.
\label{PdP2toPdP3}
} 

We see that, in addition to the $PdP3b$ and $PdP3c$ obtained as 
blowups of $dP2$ studied in the previous subsection, 
we also obtain a third new singularity which we call $PdP3a$.
As before, we again use the Inverse Algorithm to find the quiver and
superpotential. These are given below:
\beq
\hspace{-3cm}
\ba{ll}
\ba{l}\epsfxsize = 10cm \epsfbox{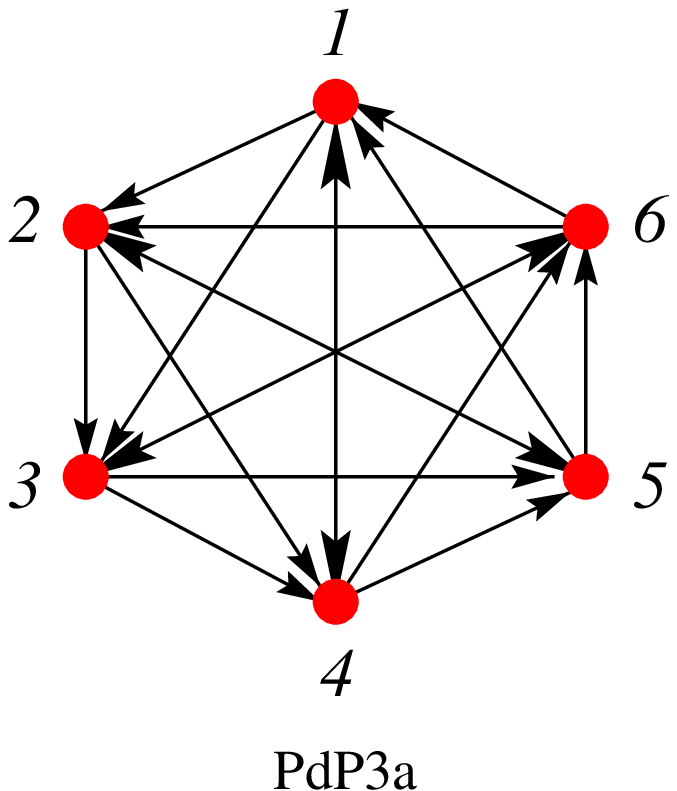}\ea & 
\hspace{-2cm}
\ba{rcl}
&& W_{PdP3a} \nn \\
&& = [ X_{13} X_{36} X_{61} + X_{24} X_{41} X_{12}+ X_{35} X_{52}
 X_{23}+ X_{46} X_{63} X_{34} \nn \\
& & \quad + X_{51} X_{14} X_{45} + X_{62} X_{25} X_{56}] -[X_{12} X_{25} X_{51}
+X_{23} X_{36} X_{62} \nonumber \\
& & \quad + X_{34} X_{41} X_{13}+ X_{45} X_{52} X_{24} + X_{56} X_{63} X_{35}
+ X_{61} X_{14} X_{46}] \ . 
\ea
\label{PdP3a}
\ea\eeq
There is only one toric phase for this theory.
Its quiver is the same as that of model I of $dP3$
if we ignore the
three bi-directional arrows connecting nodes (1,4), (2,5)
and (3,6).
The superpotential, written in the manner of \eref{PdP3a}, 
clearly manifests a $D_6$ symmetry:
a $\IZ_6$ factor which is a $60^o$ degree rotation, and a $\IZ_2$
factor exchanging  $2\leftrightarrow 6,3\leftrightarrow 5$ together
with field conjugation.

%
\section{Higgsing, Splitting and Blowing Up}

Now we have constructed a host of theories, related to each other by
blowups in the toric diagram and whose quivers can be obtained either
from the toric $(p,q)$-webs or from the Inverse Algorithm. We have
shown that different geometries may give rise to the same quiver if we
ignored bi-directional arrows which can not be encoded into the rule
\eref{interrule} for the webs anyway. We have insisted that there is a
one-one correspondence between the toric diagram and the ``toric''
$(p,q)$-web, which is the graph-dual of the toric diagram. On the
other hand, we have introduced the notion of the ``quiver''
$(p,q)$-web which is simply a web that gives the right quiver by
\eref{interrule}. Indeed, different geometries may
have the same ``quiver'' $(p,q)$-web even though their ``toric''
$(p,q)$-webs obviously differ.

To further establish the one-to-one correspondence between 
the toric data and toric $(p,q)$-web diagrams, we now show that
these theories above are indeed related to each other from various
perspectives. This is in the spirit of the (un)higgsing 
mechanism of \cite{unhiggs}. What we have used in the previous
section is sequential blow-ups (blow-downs) 
in the toric geometry corresponding to the removal (addition) of toric
points. Now we will first show that the theories can indeed be higgsed
from each other from a field theory point of view. We then show that
this is also in accordance with the splitting procedure in the
$(p,q)$-web picture.

For example,
from the toric data of $dP1$, we see that it can be embedded into
both $dP2$ and $PdP2$ as shown in \fref{dP1todP2}.
From the point of view of $(p,q)$-webs,
this process is realized by splitting  one external
leg in the  $(p,q)$-web diagram of $dP1$ \cite{Franco:2002ae}. 
To keep only one
interior point in the toric data, we get only two
inequivalent $(p,q)$-web diagrams by this splitting process.
From the perspective of the field theory on the D-brane probe, 
this means that
we can higgs the field theories of $dP2$ and $PdP2$ to that of 
$dP1$. This will be our first consistency check.


\subsection{Higgsing: Checks in the Field Theory}

We begin with checks from the field theory which lives on the D3-brane
probe world volume. We will show that the theories obtained above by
the blowups in the geometry and using the Inverse Algorithm are indeed
inter-related by the higgsing mechanism.

\subsubsection{From $PdP2$ to $dP1$}
To higgs down from $PdP2$ to $dP1$, we start from the superpotential
(\ref{WPdP2}).  By giving the field $X_{14}$ a non-zero vacuum expectation 
value, the fields $X_{42}, X_{21}$ become massive and must be integrated
out. The final theory will be left with only $13-3=10$ fields, which turns
out to be precisely those of $dP1$. Indeed, after integrating out
the massive fields and relabelling fields $X_{45}\rightarrow Y_{15},
X_{34}\rightarrow Z_{31}$, we obtain the following superpotential:
\bea
W & =
 & X_{31} X_{15} X_{53}- Y_{31} Y_{15} X_{53} -X_{31} X_{12} Y_{23}
+ Y_{31} X_{12} X_{23} \nn \\
& & \quad -X_{52} X_{23} Z_{31} X_{15} + X_{52} Y_{23} Z_{31} Y_{15} \ .
\eea
This is exactly the superpotential of $dP1$ given in \eref{summary}. 

\subsubsection{From $PdP3b$ to $dP2$}
Let us start from the phase I of $PdP3b$. First we give $X_{62}$ a non-zero
VEV. In this case, fields $X_{16}$ and
$X_{21}$ will become massive and should be integrated out, so we are
left with $16-3=13$ fields in the final theory.  After working 
out the superpotential as above, it is exactly that of
the phase I of $dP2$ given by $W_{dP2_I}$ in \eref{summary}. Next, if we 
give $X_{41}$ a non-zero VEV, fields $X_{16},X_{64}, X_{15}, X_{54}$ 
will get mass and we are left with $16-5=11$ fields. 
After integrating out the
massive fields we get exactly the superpotential of phase II of $dP2$
given by $W_{dP2_{II}}$ in \eref{summary}.

Let us now start with phase II of $PdP3b$. First, if we give field 
$X_{41}$ a VEV, it is easy to show that the final superpotential is
exactly $W_{dP2_I}$. On the other hand, if we give field $X_{26}$
a non-zero VEV, we will obtain $W_{dP2_{II}}$. 
Recall that phase II of $PdP3b$ has the same quiver as phase II of $dP3$
and both of them can higgs down to $dP2$. Therefore, by doing the reverse
procedure of unhiggsing, this gives us a non-trivial 
example that when we unhiggs one theory to a given quiver diagram, we 
may reach more than one final theory.

On the other hand, if we give $X_{25}$ the VEV, this will give the 
$PdP2$ theory
which as we have seen has same quiver as the phase II of $dP2$ except
the bi-directional arrow. Here we see
a very good example
how the $(p,q)$-web diagrams guide us which theory to higgs to which
and where ambiguities may arises due to the bi-directional arrows. 
We will discuss this issue more in the next section.

\subsubsection{Summary of Higgsings in the Field Theory}
The calculation is thus standard and we will not present the details
here for all the theories. What we will find is an intricate web of
theories inter-related by the various higgsings. 
In the table below we present which theories, in the vertical, can be
higged to which theories, in the horizontal. We explicitly indicate
which fields acquire non-zero VEV's.
\beq
{\small
\begin{array}{|c|c|c|c|} \hline
&  dP2_I  & dP2_{II}   & PdP2   \\  \hline  \hline
dP3_I  &    &  X_{12},X_{23},X_{34},X_{45},X_{56}, X_{61}  & \\  \hline
dP3_{II} &  X_{32},X_{41}  &  X_{25},X_{54},X_{64}, X_{26}  & \\  \hline
dP3_{III} &  X_{46},X_{26}  &  X_{32},X_{34},X_{24},X_{12}  & \\  \hline
dP3_{IV} &  X_{51},X_{53},X_{43},X_{42},X_{52},X_{41}  & & \\  \hline
 \hline
PdP3a & & &  X_{12},X_{23},X_{34},X_{45},X_{56},X_{61}  \\ \hline
  \hline
PdP3b_{I} &  X_{64},X_{62}  &  X_{41},X_{21}  &  X_{43},X_{23}  \\ \hline
PdP3b_{II} &  X_{32},X_{41}  &  X_{26}, X_{54}  &  X_{25},X_{64}  \\ \hline
PdP3b_{III} & &  X_{12},X_{61},X_{34},X_{45}  &  X_{23},X_{56}  \\ \hline
 \hline
PdP3c_{I} & &  X_{12},X_{61},X_{34},X_{45}  &  X_{23},X_{56}  \\ \hline
PdP3c_{II} &  X_{32},X_{41}  &  X_{26},X_{54}  &  X_{25},X_{64}  \\ \hline
  \hline
\end{array}
}
\eeq

We remark two points.
First, since the toric diagram of 
$dP2$ can not be embedded into that of $PdP3a$, they should not
be related by higgs mechanism.
This fact is shown by the empty entry in the above table. 
This same argument holds between $dP3$ and $PdP2$.
Second, starting from phase I of $dP3$ we can never reach
phase I of $dP2$ by higgsing. Conversely,
$dP3_I$ can never be reached by
the un-higgsing method of \cite{unhiggs} applied to $dP3_I$.
This indicates that the starting point of the un-higgsing process
will effect the final result.

\subsection{Splitting: Higgsing in the $(p,q)$-Web}
Now, as detailed very nicely in Section 4.1 of \cite{Franco:2002ae},
the process of (un)higgsing in the field theory, or equivalently, the
process of adding/deleting nodes in the toric diagram, has a simple
counterpart in the $(p,q)$-web picture. This simply corresponds to the
splitting and combining of external legs (dual to the addition and
deletion of points in the toric diagram). The archetypal example is
given in \fref{PdP2toF2} below.

A strong evidence for the identification of different 
$(p,q)$-webs with different phases given in \cite{Franco:2002ae} is the
ability to use the $(p,q)$-web 
to higgs very conveniently. The basic
assumption behind this idea is the premise
that the quiver diagram of a given 
$(p,q)$-web is calculated by the intersection number
a la \eref{interrule}.
As we have seen in the previous discussions, such an assumption is too
strong.
First, it is possible that there are bi-directional arrows in the
quiver diagram. 
These are not captured by the intersection number which are necessarily
antisymmetric.
Second, when there is more than one interior point in the toric
diagram, i.e., when we have parallel legs, we need more 
than one pair of $(p,q)$ charges to calculate the intersection. 
This point was shown in \cite{Feng:2002kk} in the mirror
picture\footnote{We 
thank Amer Iqbal for clarifying this point to us.}. 
Moreover, we will adhere to the following constraint.
When we read out which fields should get a non-zero VEV from the
$(p,q)$-web, we insist that only nearby external legs are combined.  
This condition is not required from the point of
view of field theories because there are a number of fields that
can develop non-zero VEVs to reach the desired result. However
from the point of view of toric
resolutions and indeed from the $(p,q)$-web, this condition is
natural.

Bearing these points in mind, we now
construct the ``quiver $(p,q)$-webs'' for the toric singularities
studied above (incidentally, these
diagrams all have only one interior point). We will reverse-engineer
the $(p,q)$-charges so that the desired quiver diagrams can be
calculated using \eref{interrule} and that the various higgsings are
in accordance with the conditions in the preceding paragraph.
Therefore, unlike the ``toric $(p,q)$-web'' which
is uniquely determined by graph dualisation once the toric diagram is
given, for the ``quiver $(p,q)$-web'' we construct the web
from the quiver diagram {\it a postiori} to show that the
correspondence may not be one-to-one.
For example, as stated earlier, the quiver $(p,q)$-web of
$PdP2$ will be identical to that of phase II of $dP2$ whereas their
toric $(p,q)$-webs clearly differ.

Indeed, we would like to emphasise that
the toric $(p,q)$-web, obtained from the geometry alone, is perhaps
a more fundamental concept. 
However, since the 
quiver $(p,q)$-web encodes the information of the quiver diagram,
it can be used to efficiently encode which field gets a non-zero
VEV as advocated in \cite{Franco:2002ae}. The 
$(p,q)$-webs identified with particular toric dual phases of a theory
in \cite{Franco:2002ae} are precisely these ``quiver $(p,q)$-webs.''
Thus armed, let us now use a few examples to demonstrate how to use the
quiver $(p,q)$-web to perform higgsing and what ambiguities may arise.

\EPSFIGURE[ht]{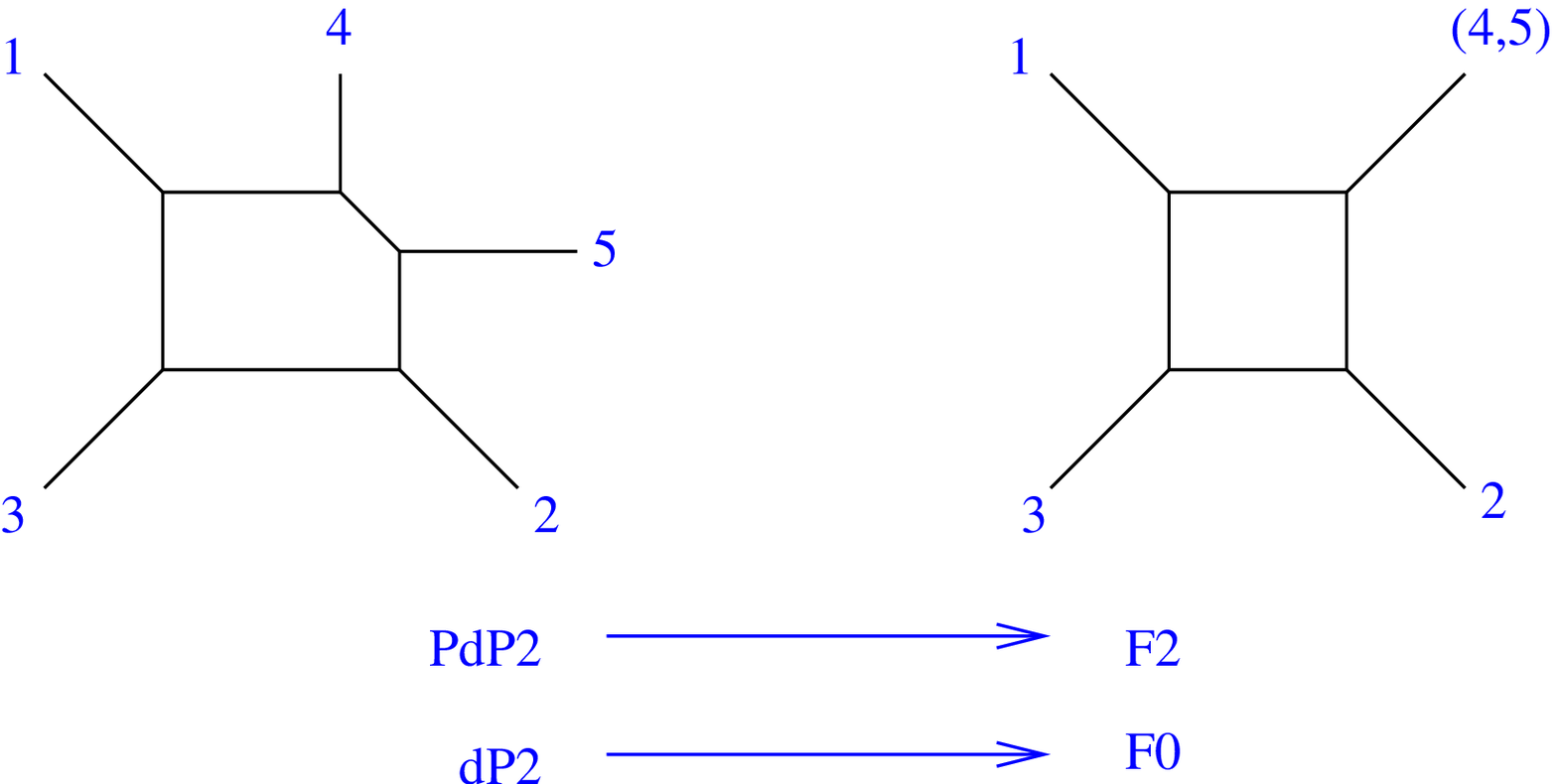,width=13cm}
{The higgsing of the quiver $(p,q)$-web of either 
$PdP2$ or $dP2_{II}$ to that of either $F_2$ or $F_0$. 
\label{PdP2toF2}
}

\subsubsection{Bi-directional Arrows}
The first example is depicted in Figure \ref{PdP2toF2}.
Since $F_2$ can be embedded into both $dP2$ and $PdP2$,
we should expect to be able to higgs $PdP2$ and $dP2$ to $F_2$
as was done in the previous sections. 
As we have seen, the
quiver diagram of $PdP2$ is same as phase II of $dP2$ up to a
bi-directional arrow, so the corresponding quiver $(p,q)$-webs
are same. From Figure \ref{PdP2toF2} it is easy
to see that by combining external legs $4$ and $5$ of left web
we can get the web on the right. 

According to the results in \cite{Franco:2002ae},
this means that if we give the field $X_{45}$ a non-zero VEV, we can 
higgs down $PdP2$ (or $dP2$) to $F_2$. We have shown this,
by using the explicit superpotential, in previous sections.
This example demonstrates that while it is very simple to
read out the higgsing information from the quiver $(p,q)$-web,
the identification of field theories to a specific quiver $(p,q)$-web 
is ambiguous.
The higgsing process dictated by a quiver $(p,q)$-web can indeed
result in 
different field theories, depending on the theory we choose to associate
with the quiver $(p,q)$-web.
Thus, the same quiver $(p,q)$-webs
can tell us the higgsing process of $dP2$ to $F_0$ or from $PdP2$ to
$F_2$.
\subsubsection{Adjacent Parallel Legs}
The above ambiguities for the quiver $(p,q)$-web is by no means
the whole story.
The procedure to read off the higgsed fields becomes much trickier
when considering $(p,q)$-webs with adjacent parallel external legs.
For example, knowing the quiver diagram of phase II of $PdP3b$, we can
draw its corresponding 
quiver $(p,q)$-web as either part (a) or (b) of \fref{PdP3bexample}.
\EPSFIGURE[ht]{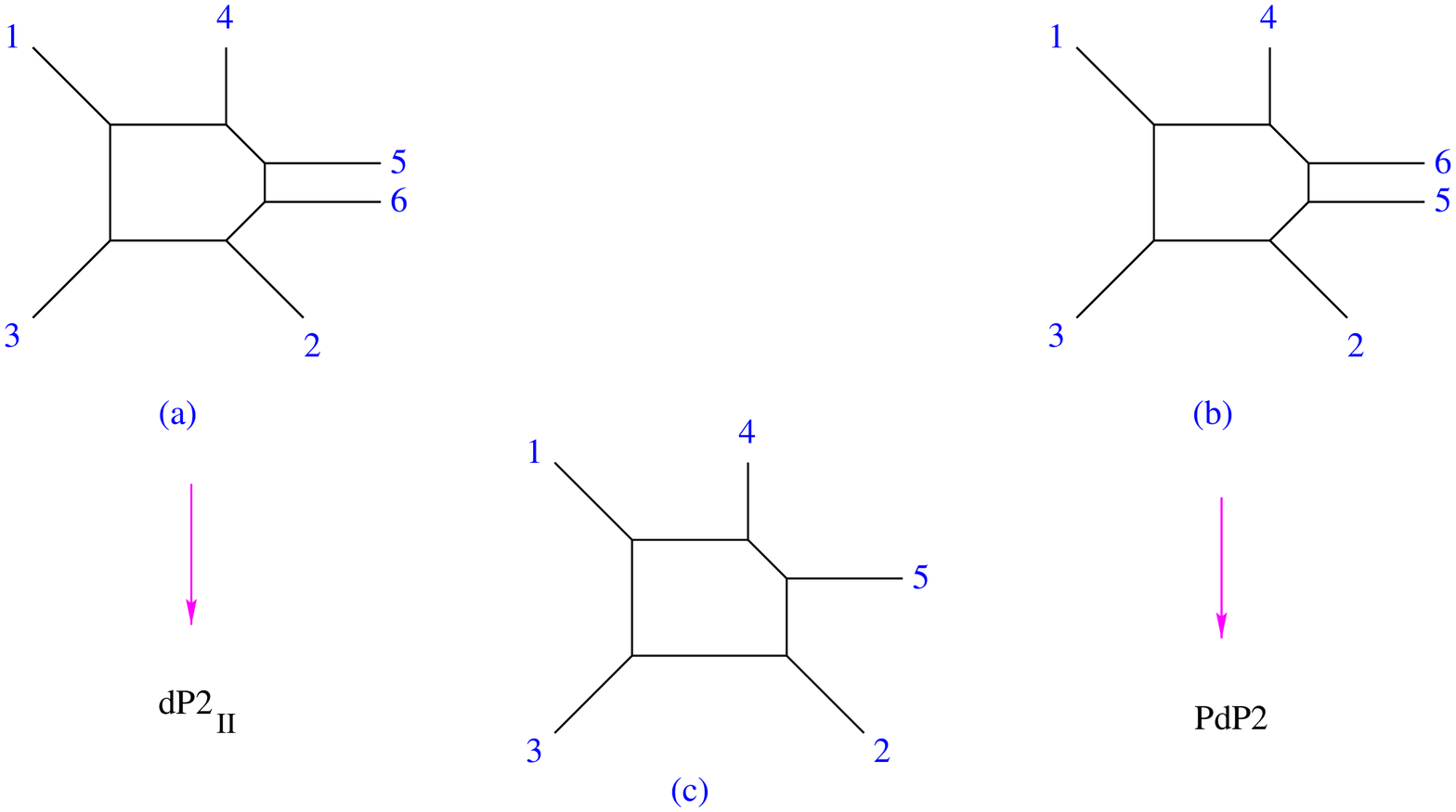,width=15cm}
{Higgsing the Phase II theory of $PdP3b$ down to $dP2$ or $PdP2$. Even
though the latter two have the same $(p,q)$-webs, their corresponding
gauge theories differ by bi-directional arrows. The 
different orders of label $5$ and $6$ (a) and (b) higgs down to these
two different theories. 
\label{PdP3bexample}
}
These quiver $(p,q)$-webs in \fref{PdP3bexample}
differ from each other only by the exchange of
legs $5$ and $6$ because these nodes are identical in the quiver
diagram.  However, when
we try to higgs these two webs down to the quiver $(p,q)$-web of $dP2$
in Part (c),  
this difference in leg ordering becomes manifest in the higgsing
procedure. Using the diagram in Part (a), by combining
legs $2$ and $6$ we reach the phase II of $dP2$, but from Part (b)   
by combining legs $2$ and $5$ we reach $PdP2$. These are, we recall,
different gauge theories.

This example 
demonstrates the tricky part of using quiver $(p,q)$-web in performing
higgsing. In fact,
this also solves the following
puzzle of higgsing down from the $(p,q)$-web of
$\IC^3/\IZ_3\times \IZ_3$. 
The requirement of combining only nearby legs tell us that there
are four and only four different $(p,q)$-webs which we can reach from 
$\IC^3/\IZ_3\times \IZ_3$ by higgsing three fields. These are
precisely the four kinds of 
toric singularities, viz., $dP3$, $PdP3a$, $PdP3b$ and $PdP3c$.
However as we saw in \sref{s:case} there are 
a total ten phases for $dP3$, $PdP3a$, $PdP3b$ and $PdP3c$.
The answer to this puzzle is that the toric $(p,q)$-web and quiver
$(p,q)$-web of $\IC^3/\IZ_3\times \IZ_3$ are identical but not 
so for the partial resolutions thereof.
By switching the positions of the external leg labels in the
$(p,q)$-web, we can obtain the total of ten phases of these four toric
singularities. Let us now see how this is done in detail.

\begin{figure}[ht]
\centerline{
$
A = \tmat{
  0 & 1 & 0 & 1 & 0 & 0 & 0 & 0 & 1 \cr 0 & 0 & 1 & 0 & 1 & 0 & 1 & 0 & 0 \cr 1
& 0 & 0 & 0 & 0 & 
   1 & 0 & 1 & 0 \cr 0 & 0 & 1 & 0 & 1 & 0 & 1 & 0 & 0 \cr 1 & 0 & 0 & 0 & 0 &
1 & 0 & 1 & 0 \cr 0 &
   1 & 0 & 1 & 0 & 0 & 0 & 0 & 1 \cr 1 & 0 & 0 & 0 & 0 & 1 & 0 & 1 & 0 \cr 0 &
1 & 0 & 1 & 0 & 0 & 
   0 & 0 & 1 \cr 0 & 0 & 1 & 0 & 1 & 0 & 1 & 0 & 0 \cr  
},
\qquad \qquad
\ba{r}
  \epsfxsize = 12cm
  \epsfbox{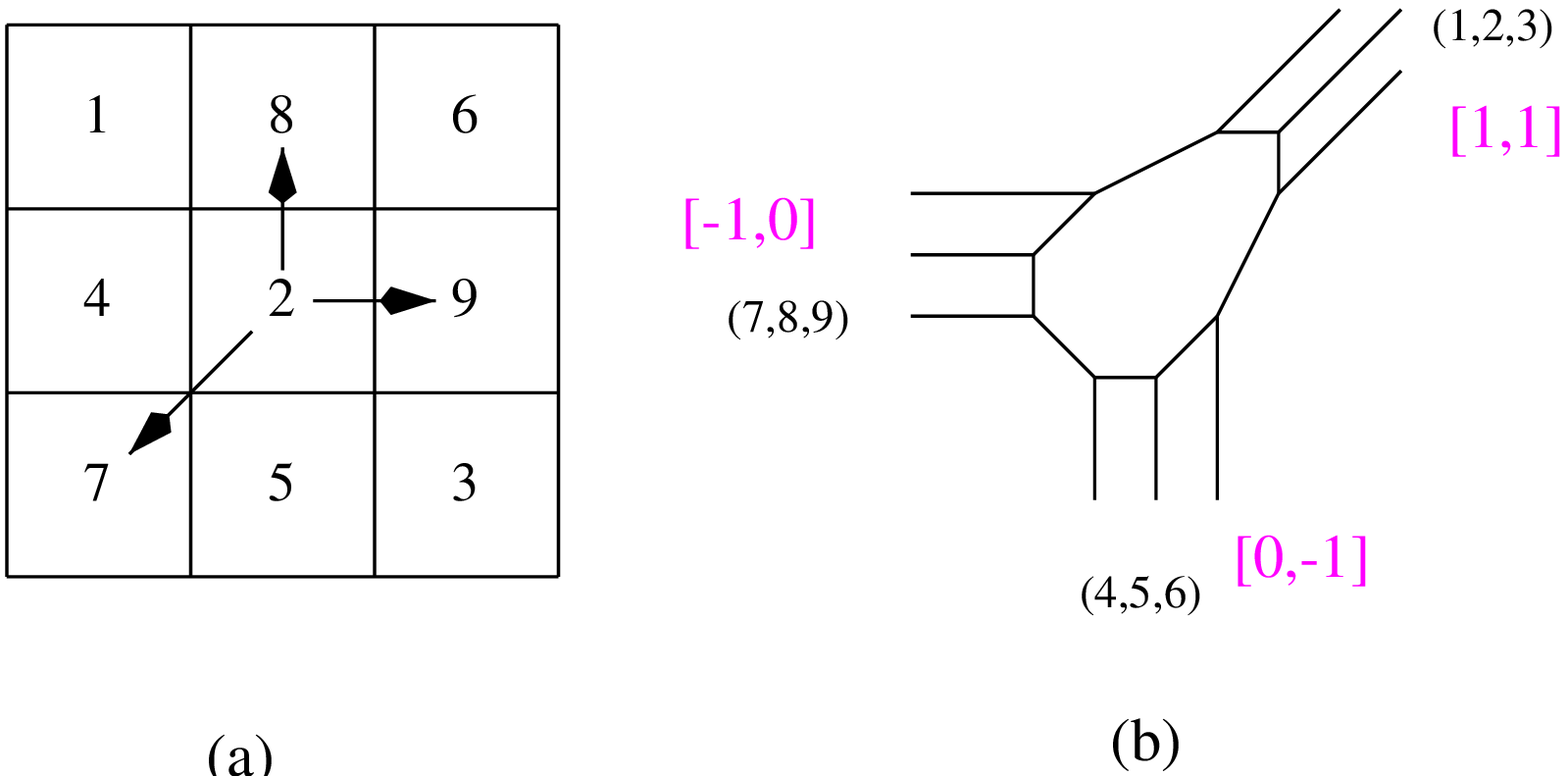}
\ea$}
  \caption{(a) The brane box model for the matter content
of $\IC^3/\IZ_3\times \IZ_3$ and (b) its
corresponding quiver-$(p,q)$-web.
Notice that we have three equivalent groups of external legs which we
have labelled as $(123)$, $(456)$ and $(789)$.
For every group, the order of nodes has not been specified, the
$(p,q)$ charges have been written in square brackets. For reference,
we have included the quiver matrix for the theory.}
\label{C3}
\end{figure}

To start, the matter content for
$\IC^3/\IZ_3\times \IZ_3$ is given in Part (a) of \fref{C3} and the
corresponding quiver $(p,q)$ is drawn in Part (b).
The superpotential of the theory is:
\bea
W & = & \phi_{18} \phi_{85} \phi_{51} - \phi_{17} \phi_{75} \phi_{51}
+\phi_{86} \phi_{63} \phi_{38} -\phi_{85} \phi_{53} \phi_{38} +
\phi_{61} \phi_{17} \phi_{76}  -\phi_{63} \phi_{37} \phi_{76} \nonumber \\
&&+\phi_{42} \phi_{28} \phi_{84}
-\phi_{41} \phi_{18}  \phi_{84}
+ \phi_{29} \phi_{96} \phi_{62} -\phi_{28} \phi_{86} \phi_{62} 
+ \phi_{94} \phi_{41} \phi_{19}-\phi_{96} \phi_{61} \phi_{19} \nonumber \\
&&+ \phi_{75} \phi_{52} \phi_{27}-\phi_{74} \phi_{42} \phi_{27}
+ \phi_{53} \phi_{39} \phi_{95}-\phi_{52} \phi_{29} \phi_{95}
+\phi_{37} \phi_{74} \phi_{43}-\phi_{39} \phi_{94} \phi_{43} \ .
\eea
We now apply the higgsing
mechanism to this orbifold theory to obtain all ten phases of 
$dP3$, $PdP3a$, $PdP3b$, and $PdP3c$.

From the corresponding toric $(p,q)$-web of $\IC^3/\IZ_3\times \IZ_3$
in \fref{C3}, we higgs down from the $U(1)^9$ theory
down to a $U(1)^6$ theory by combining adjacent legs. We consequently
arrive at four theories (cf.~\cite{Franco:2002ae}):
\beq
\ba{|c|c|c|}\hline
&&\mbox{legs combined} \\ \hline
(1) & dP3 & [-1,0]+[1,1], \ [1,1]+[0,-1], \ [0,-1]+[-1,0] \\ \hline
(2) & PdP3b & [1,1]+[0,-1], \ [0,-1]+2[-1,0] \\ \hline
(3) & PdP3c & [0,-1]+2[1,1], \ [-1,0]+[1,1] \\ \hline 
(4) & PdP3a & 3[-1,0]+[1,1] \\ \hline
\ea
\eeq
To obtain all ten theories
we simply change the ordering of external legs in Part (b) of
\fref{C3} and hence alter the fields which obtain non-zero VEV. 
We summarise the relevant permutations below. The notation is such
that $(a_1a_2)(b_1b_2)(c_1c_2)$ means that the pairs of legs in each
of the three brackets are to be combined.
\beq
\epsfxsize = 17cm\epsfbox{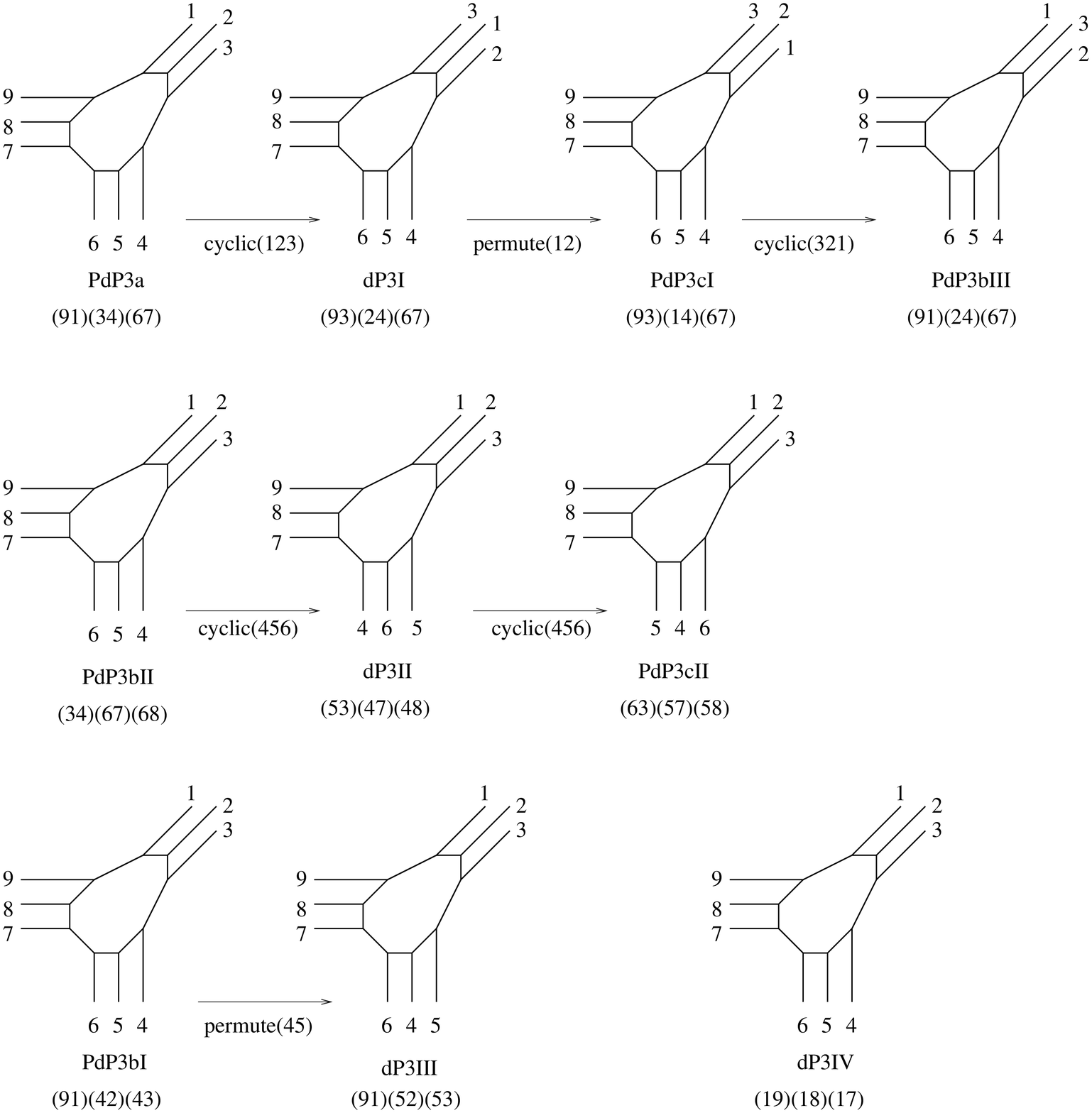}
\eeq

In summary, the reasons that quiver $(p,q)$-web can describe more
than one gauge theories are (1) it does not contain information
about the bi-directions bifundamental fields; (2) the ordering
ambiguities of parallel lines affects the Higgsing process. 
Therefore, we do not take quiver $(p,q)$-web as
a fundamental concept, though it is evidently a very useful 
tool.

\newpage

%
%
\section*{Acknowledgements}
We would 
like to extend our sincere gratitude to A.~Hanany, who, still like
a father and a friend, has taken the pains to review the manuscript
and to offer insightful comments. We are indebted to S.~Franco
and A.~Iqbal for many enjoyable discussions. YHH also thanks S.~Kirbos
for charming diversions.

The research of BF is funded in part by the SNS of the IAS and
an NSF grant PHY-0070928, and that of YHH, by the Dept of Physics, UPenn,
a U.S.~DOE Grant $\#$DE-FG02-95ER40893 as well as
an NSF Focused Research Grant
DMS0139799 for ``The Geometry of Superstrings.'' FHL is
also grateful to the Deutsche Bank of New York.

\vspace{2in}

\bibliographystyle{JHEP}

\end{document}